\documentclass{article}

\usepackage{PRIMEarxiv}

\usepackage[utf8]{inputenc} % allow utf-8 input
\usepackage[T1]{fontenc}    % use 8-bit T1 fonts
\usepackage{hyperref}       % hyperlinks
\usepackage{url}            % simple URL typesetting
\usepackage{booktabs}       % professional-quality tables
\usepackage{amsmath}
\usepackage{amsthm}
\usepackage{amssymb}
\usepackage{amsfonts}
\usepackage{mathtools}
\usepackage{amsfonts}       % blackboard math symbols
\usepackage{nicefrac}       % compact symbols for 1/2, etc.
\usepackage{microtype}      % microtypography
\usepackage{fancyhdr}       % header
\usepackage{graphicx}       % graphics
\graphicspath{{media/}}     % organize your images and other figures under
\usepackage{times}
\usepackage{subfig}
\usepackage{xcolor}
\usepackage{array}
\usepackage{cite}
\usepackage{multirow}
\usepackage{adjustbox}
\usepackage[scr=dutchcal]{mathalfa}
\usepackage[font=small,labelfont=bf]{caption}
\usepackage{algorithm}
\usepackage{algpseudocode}
\usepackage{enumitem}

\DeclareMathOperator{\diag}{diag}
\DeclareMathOperator{\tr}{tr}

\newtheorem{theorem}{Theorem}
\newtheorem{lemma}{Lemma}
\newtheorem{proposition}[theorem]{Proposition}
\newtheorem{corollary}{Corollary}

\newtheorem{assumption}{Assumption}

\newtheorem{remark}{Remark}

\usepackage{tikz-network}
\tikzset{
    block/.style = {draw, fill=white, rectangle, minimum height=.75cm, minimum width=1.25cm},
    tmp/.style  = {coordinate}, 
    sum/.style= {draw, fill=white, circle},
    input/.style = {coordinate},
    output/.style= {coordinate},
    pinstyle/.style = {pin edge={to-,thin,black}
    }
}

\title{Distributed Adaptive Estimation with \\ ISS Guarantees for Sensor Networks with \\ Partially Unknown Source Dynamics
}

\author{
  Moh Kamalul Wafi, Hamidreza Montazeri Hedesh, and Milad Siami\\
  Department of Electrical \& Computer Engineering \\
  Northeastern University \\
  Boston, MA 02115 USA \\
  \texttt{\{wafi.m, montazerihedesh.h, m.siami\}@northeastern.edu} \\
}

\begin{document}
\maketitle

\begin{abstract}
This paper studies distributed adaptive estimation over sensor networks with partially unknown source dynamics. We present parallel continuous-time and discrete-time designs in which each node runs a local adaptive observer and exchanges information over a directed graph. For both time scales, we establish stability of the network coupling operators, prove boundedness of all internal signals, and show convergence of each node’s estimate to the source despite model uncertainty and disturbances. We further derive input-to-state stability (ISS) bounds that quantify robustness to bounded process noise. A key distinction is that the discrete-time design uses constant adaptive gains and per-step regressor normalization to handle sampling effects, whereas the continuous-time design does not. A unified Lyapunov framework links local observer dynamics with graph topology. Simulations on star, cyclic, and path networks corroborate the analysis, demonstrating accurate tracking, robustness, and scalability with the number of sensing nodes.
\end{abstract}


\allowdisplaybreaks

\section{Introduction}
Distributed state estimation over sensor networks has emerged as an underlying problem in multi-agent systems, with applications spanning environmental monitoring, surveillance systems, distributed robotics, and cooperative control \cite{R1,R2}. In these scenarios, a network of sensing agents must collectively estimate the state of a target (source) whose dynamics may be known or only partially known. The distributed nature of the problem necessitates algorithms that coordinate local observations and computations over networks.

Existing approaches to distributed state estimation broadly fall into consensus-based observers and distributed Kalman filtering. Consensus-based methods \cite{R4,R5,R6} achieve agreement among agents through iterative averaging but typically assume the source dynamics are perfectly known. Distributed Kalman filters \cite{R7,R8,R9} handle stochastic uncertainty but require knowledge of noise statistics and system matrices. Recent extensions \cite{R10} address time-varying topologies and communication constraints, yet a fundamental assumption persists: the source dynamics matrices are known a priori.

In many practical scenarios, however, the source dynamics contain parametric uncertainty—for instance, an environmental process with unknown reaction rates, or a vehicle with uncertain mass and damping coefficients. Recent works have explored distributed adaptive estimation to address such uncertainties, combining local learning with networked coordination \cite{R11,R13,R14}.
While these studies achieve asymptotic convergence under ideal or noise-free conditions, they typically neglect bounded process disturbances and do not establish input-to-state stability (ISS). 

This motivates us to study adaptive distributed estimation in directed sensor networks where each node must simultaneously estimate the source state and identify unknown dynamics while coordinating only with in-neighbors. This paper develops continuous- and discrete-time frameworks for such settings, providing rigorous input-to-state stability (ISS) guarantees via Lyapunov analysis and linear matrix inequality (LMI) conditions, and explicitly revealing how network coupling, adaptation dynamics, and disturbance robustness interact.

\textbf{Our main contributions are:}
\begin{enumerate}[leftmargin=*]
    \item A unified network-level formulation (via Kronecker structures) that decouples local observer design from graph topology, enabling modular stability analysis;
    
    \item For continuous-time systems: smooth gradient-based adaptive laws whose closed-loop error converges under a spectral/LMI condition on the coupling operator~$\mathcal H$, together with an ISS bound in the presence of bounded source disturbances;

    \item For discrete-time systems: a gradient-based adaptive law analyzed under analogous (Point 2) spectral/LMI conditions on the coupling operator $\mathcal{S}$, but augmented with regressor normalization and a constrained adaptation gain (within a stability range) to address sampling and non–strongly convex regressors. The design guarantees bounded parameter iterates, error convergence, as well as an ISS bound with respect to bounded source disturbances.
    
    \item For both time scales, explicit trade-offs are established among network connectivity, local damping, and source growth, including robustness to norm-bounded uncertainties; providing transparent design guidelines;
    
    \item Simulation studies on star, cyclic, and path graphs (and larger networks) that validate tracking performance, ISS robustness, and scalability with increasing numbers of sensing nodes.
\end{enumerate}

% A key distinction between the two formulations lies in the adaptation mechanics: in continuous time, the gradient update evolves smoothly and admits a Lyapunov analysis without regressor normalization; in discrete time, sampling and non–strongly convex regressors necessitate per-step normalization and gains constrained to a stability range (rather than arbitrary positive constants) to prevent parameter blow-up.

%===============================================================================
\textbf{Notations.} 
$I_p$ is the $p\times p$ identity, and $\diag\{A_i\}\in\mathbb{R}^{pn\times pm}$ for $i=1,\dots,p$ forms a block-diagonal matrix with blocks $A_i\in\mathbb{R}^{n\times m}$. Vectors of ones and zeros in $\mathbb{R}^p$ are $\mathbf{1}_p$ and $\mathbf{0}_p$. The Kronecker product is $A\otimes B$, $\tr(A)$ is the trace, $\lambda(A)$ denotes eigenvalues.
% For matrix $A$, $\|A\|_2$ denotes the induced 2-norm and $\|A\|_F$ the Frobenius norm. For symmetric $A$, $A\succ 0$ ($A\succeq 0$) means positive (semi-)definite. For vector $v$, $\|v\|$ is the Euclidean norm.

\section{Communication Network}\label{sec:ComNetwork}

We consider a network of $m{+}1$ agents indexed by $\mathcal{V}=\{0,1,\dots,m\}$, where agent 0 is the source (system being estimated) and agents $1,\dots,m$ are sensing nodes that run local adaptive observers and exchange information.

Communication is modeled by a weighted directed graph \(\mathcal{G}=(\mathcal{V},\mathcal{E},\mathcal{W})\) with edge set \(\mathcal{E}\subseteq\mathcal{V}\times\mathcal{V}\) and nonnegative weights \(\mathcal{W}=\{w_{ij}\}\). An edge \((i,j)\in\mathcal{E}\) means agent \(i\) receives information from agent \(j\) with weight \(w_{ij}>0\); otherwise \(w_{ij}=0\). The in-neighborhood of agent \(i\) is \(\mathcal{N}_i \coloneqq \{\,j\in\mathcal{V}\mid(i,j)\in\mathcal{E}\,\}\). The weights \(\{w_{ij}\}\) are algorithmic design coefficients (not physical signal strengths); how much weight agent $i$ places on information from agent $j$.

\begin{figure}[t!]
    \centering
    \scalebox{0.90}{{\begin{tikzpicture}
            \centering
            % \draw[black] (-1.5,-1.5) rectangle (1.5,1.5);
            \Text[x=0,y=1.3,fontsize=\small]{$\mathcal{G}$};
            \Vertex[x=-.375,y=.375,label=$0$,color=red,opacity=0.1,size=.5]{L}
            \Vertex[x=1,y=1,label=$1$,color=green,opacity=0.1,size=.5]{1}
            \Vertex[x=-1,y=1,label=$2$,color=green,opacity=0.1,size=.5]{2}
            \Vertex[x=-1,y=-1,label=$3$,color=green,opacity=0.1,size=.5]{3}
            \Vertex[x=0,y=-1,label=$4$,color=green,opacity=0.1,size=.5]{4}
            \Vertex[x=1,y=0,label=$5$,color=green,opacity=0.1,size=.5]{5}
            \Edge[Direct,color=red,label=$w_{10}$](L)(1)
            \Edge[Direct,color=red,label=$w_{40}$](L)(4)
            \Edge[Direct,color=red,label=$w_{50}$](L)(5)
            \Edge[Direct,label=$w_{21}$](1)(2)
            \Edge[Direct,label=$w_{32}$](2)(3)
            \Edge[Direct,bend=30,label=$w_{45}$](5)(4)
            \Edge[Direct,bend=30,label=$w_{54}$](4)(5)
        \end{tikzpicture}}}
    \qquad
    \scalebox{0.90}{{\begin{tikzpicture}
            \centering
            % \draw[black] (-1.5,-1.5) rectangle (1.5,1.5);
            \Text[x=0,y=1.3,fontsize=\small]{$\mathcal{G}_m$};
            \Vertex[x=1,y=1,label=$1$,color=green,opacity=0.1,size=.5]{1}
            \Vertex[x=-1,y=1,label=$2$,color=green,opacity=0.1,size=.5]{2}
            \Vertex[x=-1,y=-1,label=$3$,color=green,opacity=0.1,size=.5]{3}
            \Vertex[x=0,y=-1,label=$4$,color=green,opacity=0.1,size=.5]{4}
            \Vertex[x=1,y=0,label=$5$,color=green,opacity=0.1,size=.5]{5}
            \Edge[Direct,label=$w_{21}$](1)(2)
            \Edge[Direct,label=$w_{32}$](2)(3)
            \Edge[Direct,bend=30,label=$w_{45}$](5)(4)
            \Edge[Direct,bend=30,label=$w_{54}$](4)(5)
        \end{tikzpicture}}}
    \qquad
    \scalebox{0.90}{{\begin{tikzpicture}
            \centering
            % \draw[black] (-1.5,-1.5) rectangle (1.5,1.5);
            \Text[x=0,y=1.3,fontsize=\small]{$\mathcal{G}_0$};
            \Vertex[x=-.375,y=.375,label=$0$,color=red,opacity=0.1,size=.5]{L}
            \Vertex[x=1,y=1,label=$1$,color=green,opacity=0.1,size=.5]{1}
            \Vertex[x=0,y=-1,label=$4$,color=green,opacity=0.1,size=.5]{4}
            \Vertex[x=1,y=0,label=$5$,color=green,opacity=0.1,size=.5]{5}
            \Edge[Direct,color=red,label=$w_{10}$](L)(1)
            \Edge[Direct,color=red,label=$w_{40}$](L)(4)
            \Edge[Direct,color=red,label=$w_{50}$](L)(5)
        \end{tikzpicture}}}
    \caption{Example of a graph $\mathcal{G}$ with $m=5$, sensing-to-sensing subgraph $\mathcal{G}_m$, and  source-to-sensing subgraph $\mathcal{G}_0$, showing the decoupling and assignment of $w_{ij}$.}
    \label{Fig:network}
\end{figure}
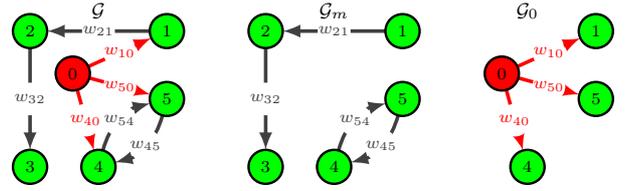

For analysis, we partition $\mathcal{G}$ into two induced subgraphs:
(i) the sensing-only subgraph $\mathcal{G}_m=(\mathcal{V}_m,\mathcal{E}_m,\mathcal{W}_m)$ with $\mathcal{V}_m=\{1,\dots,m\}$;
(ii) the source-to-sensing subgraph $\mathcal{G}_0=(\mathcal{V}_0,\mathcal{E}_0,\mathcal{W}_0)$ with $\mathcal{V}_0=\{0\}\cup\{i\in\mathcal{V}_m:(i,0)\in\mathcal{E}\}$.
Thus, $\mathcal{E}_m$ contains sensing-to-sensing links and $\mathcal{E}_0$ contains source-to-sensing links (see Fig.~\ref{Fig:network}).

For \(\mathcal{G}_m\), define the in-degree $d_i \coloneqq \sum_{j:(i,j)\in\mathcal{E}_m} w_{ij}$, $i\in\mathcal{V}_m$ and \(\mathbb{D}_m=\diag\{d_1,\dots,d_m\}\).
The adjacency and Laplacian of the sensing subgraph are $[\mathbb{A}_m]_{ij} = w_{ij}$ if $(i,j)\in\mathcal{E}_m$ (with 0 otherwise) and $\mathbb{L}_m \coloneqq \mathbb{D}_m - \mathbb{A}_m$.
For \(\mathcal{G}_0\), source-to-sensing connections are encoded by a diagonal matrix $\mathbb{A}_0 \coloneqq \diag\{w_{10},\dots,w_{m0}\}$ where $[\mathbb{A}_0]_{ii} = w_{i0}$ if $(i,0)\in\mathcal{E}_0$ and 0 otherwise.

Let $\mathbb{W}\coloneqq \diag\{w_1,\dots,w_m\}$ with $w_i\coloneqq d_i + w_{i0}$ (total incoming weight). The augmented Laplacian is
\begin{equation*}
    \mathbb{L} \coloneqq \mathbb{L}_m + \mathbb{A}_0 = \mathbb{W} - \mathbb{A}_m.
\end{equation*}
We propose the \emph{balanced} condition \(\mathbb{W}=I_m\), or $d_i + w_{i0} = 1$ $(\sum_{j\in\mathcal{N}_i} w_{ij} = 1), \forall i\in\mathcal{V}_m$. Under this condition\footnote{If \(\mathbb{W}\neq I_m\), normalization via \(\tilde w_{ij}\coloneqq w_{ij}/w_i\) for \((i,j)\in\mathcal{E}_m\) and \(\tilde w_{i0}\coloneqq w_{i0}/w_i\) yields \(\tilde{\mathbb{W}}=I_m\) and preserves \eqref{eq:ComNet:balance}; this is not required for implementation but simplifies stability analysis.},
\begin{equation}\label{eq:ComNet:balance}
    (\mathbb{L}-\mathbb{A}_0)\mathbf{1}_m=\mathbf{0}_m
    \quad\Leftrightarrow\quad
    (\mathbb{A}_m+\mathbb{A}_0)\mathbf{1}_m=\mathbf{1}_m.
\end{equation}

\begin{remark}[Source Reachability]\label{rem:reachability}
We assume: (i) at least one sensing node directly receives information from the source ($w_{i0}>0$ for some $i$), and (ii) every sensing node is reachable from the source via a directed path in $\mathcal{G}$. Under these assumptions and the balanced condition, the augmented Laplacian $\mathbb{L}$ is \emph{positive stable}: all eigenvalues have strictly positive real parts, $\Re\{\lambda_i(\mathbb{L})\}>0$ for all $i$.
\end{remark}

This property ensures that $\mathbb{L}$ admits a Lyapunov characterization essential for the input-to-state (ISS) analysis.

\section{Problem Formulation}\label{sec:ProblemFormulation}

We study distributed estimation where:
\begin{itemize}[leftmargin=*]
    \item Source (agent 0) has uncertain linear dynamics—system matrices contain unknown but bounded perturbations,
    \item Each sensing node runs local adaptive observer, exchanges estimates with in-neighbors,
    \item Not all nodes sense source directly; information propagates through $\mathcal{G}$.
\end{itemize}

We consider both continuous-time and discrete-time cases, as there are fundamental differences in constructing adaptive laws, including constant-gain adaptation and regressor normalization (due to not strongly convex cost function). Moreover, we assume \emph{full-state output} in both time scales (continuous and discrete) for clarity. This is realistic for applications like GPS tracking where measured position is the state.

\subsection{Problem 1: Unknown Continuous-time Dynamics}
The unknown continuous-time source evolves as
\begin{subequations}
\begin{align}
    \dot{x}_{0}(t) &= F_0 x_{0}(t) + G_0 v_0(t) + d_{0}(t), \label{eq:ProbFor:con:source} \\
    y_{0}(t) &= x_{0}(t) , \label{eq:ProbFor:con:output}
\end{align}
\end{subequations}
where \(x_{0}(t)\in\mathbb{R}^{n}\) is the state, \(v_0(t)\in\mathbb{R}^{p}\) is the reference input and \(y_{0}(t)\in\mathbb{R}^{n}\) is the output. The process noise \(d_{0}(t)\in\mathbb{R}^{n}\) is bounded as \(\|d_{0}(t)\|\le d^{\ast}\) for some constant \(d^{\ast}>0\). The matrices $F_{0}$ and $G_{0}$ are \emph{unknown}, defined as
\begin{equation*}
    F_{0} \coloneqq F_\ast + \Delta F\in\mathbb{R}^{n\times n}, \quad G_{0} \coloneqq G_\ast + \Delta G \in\mathbb{R}^{n\times p}.
\end{equation*}
The nominal parts $F_\ast$ and $G_\ast$ are known; the perturbations $\Delta F$ and $\Delta G$ are unknown but bounded, $\|\Delta F\|_2\le f$ and $\|\Delta G\|_2\le g$, for some constants $f,g>0$.

Sensing node $i\in\{1,\dots,m\}$ implements
\begin{subequations}
\begin{align}
    \dot{x}_{i}(t) &= H_i x_{i}(t) + (\hat{F}_{i}(t) - H_i) z_{i}(t) + \hat{G}_{i}(t) v_0(t), \label{eq:ProbFor:con:sensor} \\
    z_{i}(t) &= \sum\nolimits_{j\in\mathcal{N}_{i}} w_{ij}x_{j}(t) + w_{i0}y_{0}(t), \quad \mathcal{N}_{i}\subseteq\mathcal{V}, \label{eq:ProbFor:con:z}
\end{align}
\end{subequations}
where $x_i(t)\in\mathbb{R}^n$ is the $i$-th sensing node estimate. Both $\hat{F}_{i}(t)\in\mathbb{R}^{n\times n}$ and $\hat{G}_{i}(t)\in\mathbb{R}^{n\times p}$ are adaptive estimates of $F_0$ and $G_0$, respectively, and $H_i\in\mathbb{R}^{n\times n}$ is user-designed Hurwitz matrix. Here, $z_i(t)$ is weighted aggregate of in-neighbor estimates plus (if applicable) source output.

\textbf{Main objective}: Design a distributed adaptive law such that, under the balanced condition \eqref{eq:ComNet:balance} and the source-reachability assumption in Remark~\ref{rem:reachability}, all sensing node estimates are ISS stable with respect to the bounded disturbance on the source, i.e.,
\begin{subequations}
\begin{equation}\label{eq:ProbFor:con:ISS}
    \limsup\nolimits_{t\to\infty}\|x_i(t)-z_i(t)\| \le c_{\mathrm c}d^{\ast}, \quad \forall i,
\end{equation}
for $c_{\mathrm c} > 0$ and, in the disturbance-free case $d^\ast=0$
\begin{equation}\label{eq:ProbFor:con:goal:ISS}
    \lim\nolimits_{t\to\infty} e_i(t) \coloneqq x_i(t)-z_i(t) = 0.
\end{equation}
\end{subequations}

\subsection{Problem 2: Uncertain Discrete-time Dynamics}
To avoid confusion with the continuous-time setting, we adopt separate notation in the discrete-time case. The unknown discrete-time source evolves as
\begin{subequations}
\begin{align}
    x_{0,k+1} &= A_0 x_{0,k} + B_0 u_{0,k} + \delta_{0,k}, \label{eq:ProbFor:dis:source} \\
    y_{0,k} &= x_{0,k}, \label{eq:ProbFor:dis:output}
\end{align}
\end{subequations}
where $x_{0,k}\in\mathbb{R}^n$ defines the source state while $u_{0,k}\in\mathbb{R}^p$ and $y_{0,k}\in\mathbb{R}^n$ denote the reference input and the source output. The term $\delta_{0,k}\in\mathbb{R}^n$ represents process noise, bounded as $\|\delta_{0,k}\|\le \delta^\ast$ for some constant $\delta^\ast > 0$. The matrices $A_{0}$ and $B_{0}$ are \emph{unknown}, denoted as
\begin{equation*}
    A_0 \coloneqq A_\ast + \Delta A\in\mathbb R^{n\times n}, \quad B_0\coloneqq B_\ast + \Delta B\in\mathbb R^{n\times p},
\end{equation*}
yet the pair $(A_\ast, B_\ast)$ is known with bounds $\|\Delta A\|_2\le a$, $\|\Delta B\|_2\le b$, for some constants $a,b>0$.

Next, sensing node $i\in\{1,\dots,m\}$ evolves as
\begin{subequations}
\begin{align}
    x_{i,k+1} &= S_i x_{i,k} + (\hat{A}_{i,k} - S_i) z_{i,k} + \hat{B}_{i,k} u_{0,k}, \label{eq:ProbFor:dis:sensor} \\
    z_{i,k} &= \sum\nolimits_{j\in\mathcal{N}_{i}} w_{ij}x_{j,k} + w_{i0}y_{0,k}, \quad \mathcal{N}_{i}\subseteq\mathcal{V}, \label{eq:ProbFor:dis:z}
\end{align}
\end{subequations}
where $x_{i,k}\in\mathbb{R}^n$ is the $i$-th sensing node estimate.
The terms $\hat{A}_{i,k}\in\mathbb R^{n\times n}$ and $\hat{B}_{i,k}\in\mathbb R^{n\times p}$ are the estimates of the pair $(A_0,B_0)$ with user-designed Schur-stable matrix $S_i\in\mathbb R^{n\times n}$. Finally, $z_{i,k}$ is weighted aggregate of in-neighbor estimates (including the source if $0\in\mathcal N_i$).

\textbf{Main objective}: Design a distributed adaptive law such that, under \eqref{eq:ComNet:balance} and Remark~\ref{rem:reachability}, all sensing node estimates are ISS stable with respect to the bounded disturbance on the source, i.e.,
\begin{subequations}
\begin{equation}\label{eq:ProbFor:dis:ISS}
    \limsup\nolimits_{k\to\infty} \|x_{i,k} - z_{i,k}\| \le c_{\mathrm d}\delta^{\ast} \quad \forall i,
\end{equation}
for $c_{\mathrm d} > 0$ and, in the disturbance-free case $\delta^\ast = 0$
\begin{equation}\label{eq:ProbFor:dis:goal:ISS}
    \lim\nolimits_{k\to\infty} e_{i,k} \coloneqq x_{i,k}-z_{i,k} = 0.
\end{equation}
\end{subequations}

\subsection{Unified Notation}

To avoid repetition, we define the shorthand notation for the stacked variables (cf. \eqref{eq:ProbFor:con:source}–\eqref{eq:ProbFor:dis:goal:ISS}) used in the following sections. For any source signal $\beta_{0}(t),\beta_{0,k}\in\mathbb{R}^{q}$,
\begin{subequations}
\begin{align}
    \Bar{\beta}_{0}(t) &= \mathbf{1}_m \otimes \beta_{0}(t) \in \mathbb{R}^{mq}, 
    \label{eq:ProbFor:beta0:t} \\
    \Bar{\beta}_{0,k} &= \mathbf{1}_m \otimes \beta_{0,k} \in \mathbb{R}^{mq}.
    \label{eq:ProbFor:beta0:k}
\end{align}
\end{subequations}
For any sensing-node signal $\beta_{i}(t), \beta_{i,k} \in\mathbb{R}^{q}$,
\begin{subequations}
\begin{align}
    \Bar{\beta}(t) &= [\beta_{1}^\top(t),\dots,\beta_{m}^\top(t)]^\top \in \mathbb{R}^{mq}, 
    \label{eq:ProbFor:betai:t} \\
    \Bar{\beta}_{k} &= [\beta_{1,k}^\top,\dots,\beta_{m,k}^\top]^\top \in \mathbb{R}^{mq}. 
    \label{eq:ProbFor:betai:k}
\end{align}
\end{subequations}
For matrices $E_0, E_i, \hat{E}_{i}(t), \hat{E}_{i,k}\in\mathbb{R}^{q_1\times q_2}$,
\begin{subequations}
\begin{align}
    \mathbf{E}_0 &= I_m \otimes E_0 \in\mathbb{R}^{mq_1\times mq_2}, \label{eq:ProbFor:mat0}\\
    \mathbf{E} &= \diag\{E_1,\dots,E_m\}\in\mathbb{R}^{mq_1\times mq_2}, \label{eq:ProbFor:mati}\\
    \hat{\mathbf{E}}(t) &= \diag\{\hat{E}_{1}(t),\dots,\hat{E}_{m}(t)\}\in\mathbb{R}^{mq_1\times mq_2}, \label{eq:ProbFor:mati:t}\\
    \hat{\mathbf{E}}_k &= \diag\{\hat{E}_{1,k},\dots,\hat{E}_{m,k}\}\in\mathbb{R}^{mq_1\times mq_2}. \label{eq:ProbFor:mati:k}
\end{align}
\end{subequations}
With these definitions, we are ready to study the adaptive laws and their stability in both time scales.

\section{Adaptive Estimation for Problem 1}\label{sec:AdpEst:Problem1}

\subsection{Error Dynamics in Networked Systems}\label{subsec:Problem1:1}
In network-level form, the source evolves as
\begin{subequations}
\begin{align}
    \dot{\Bar{x}}_{0}(t) &= \mathbf{F}_0 \Bar{x}_{0}(t) + \mathbf{G}_0 \Bar{v}_0(t) +  \Bar{d}_0(t), \label{eq:Problem1:con:source} \\
    \Bar{y}_{0}(t) &= \Bar{x}_{0}(t), \label{eq:Problem1:con:output}
\end{align}
\end{subequations}
while the sensing nodes obey
\begin{subequations}
\begin{align}
    \dot{\Bar{x}}(t) &= \mathbf{H} \Bar{x}(t) + (\hat{\mathbf{F}}(t) - \mathbf{H}) \Bar{z}(t) + \hat{\mathbf{G}}(t) \Bar{v}_0(t), \label{eq:Problem1:con:sensor} \\
    \Bar{z}(t) &= (\mathbb{A}_m\otimes I_n)\Bar{x}(t) + (\mathbb{A}_0\otimes I_n)\Bar{y}_0(t). \label{eq:Problem1:con:z}
\end{align}
\end{subequations}
Here, the signals $\Bar{x}_0(t)$, $\Bar{v}_0(t)$, $\Bar{d}_0(t)$, and $\Bar{y}_0(t)$ follow \eqref{eq:ProbFor:beta0:t} whereas $\Bar{x}(t)$ and $\Bar{z}(t)$ follow \eqref{eq:ProbFor:betai:t}. The matrices $\mathbf{F}_0$ and $\mathbf{G}_0$ are expressed as in \eqref{eq:ProbFor:mat0} while the matrix $\mathbf{H}$ is denoted as in \eqref{eq:ProbFor:mati}. Also, both $\hat{\mathbf{F}}(t)$ and $\hat{\mathbf{G}}(t)$ are defined as in \eqref{eq:ProbFor:mati:t}.

Before proceeding, we denote the following assumption.
\begin{assumption}\label{assmpt:con:optimal}
    There exist optimal parameter values of the source dynamics in \eqref{eq:Problem1:con:source}, denoted as $\hat{\mathbf{F}}^\ast$ and $\hat{\mathbf{G}}^\ast$, such that $\hat{\mathbf{F}}^\ast \coloneqq \mathbf{F}_0$ and $\hat{\mathbf{G}}^\ast \coloneqq \mathbf{G}_0$. 
\end{assumption}

Then, we introduce the parameter errors
\begin{align}
    \Phi_{\mathrm{c}}(t) \triangleq \hat{\mathbf{F}}(t) - \mathbf{F}_0, \quad
    \Psi_{\mathrm{c}}(t) \triangleq \hat{\mathbf{G}}(t) - \mathbf{G}_0. \label{eq:Problem1:par_error}
\end{align}
From \eqref{eq:ProbFor:con:goal:ISS}, the estimation error defined as in \eqref{eq:ProbFor:betai:t} is 
\begin{equation}\label{eq:Problem1:con:error}
    \Bar{e}(t) = (\mathbb{L}\otimes I_n)\Bar{x}(t) - (\mathbb{A}_0\otimes I_n)\Bar{y}_0(t) \coloneqq \Bar{x}(t) - \Bar{z}(t).   
\end{equation}
Thus, the estimation error dynamics inspired by \cite{Wafi-MRAC,Wafi-LCSS24} become 
\begin{align}
    \dot{\Bar{e}}(t) &= (\mathbb{L}\otimes I_n)\dot{\Bar{x}}(t) - (\mathbb{A}_0\otimes I_n)\dot{\Bar{x}}_0(t) \nonumber\\
    &= (\mathbb{L}\otimes I_n)\bigl[\mathbf{H}\Bar{x}(t) + (\hat{\mathbf{F}}(t) - \mathbf{H})\Bar{z}(t) + \hat{\mathbf{G}}(t)\Bar{v}_0(t)\bigr] \nonumber\\
    &\quad - (\mathbb{A}_0\otimes I_n)\bigl[\mathbf{F}_0\Bar{x}_0(t) + \mathbf{G}_0\Bar{v}_0(t) + \Bar{d}_0(t)\bigr] \nonumber\\
    &= (\mathbb{L}\otimes I_n)\bigl[\mathbf{H}\Bar{x}(t) + \hat{\mathbf{F}}(t)\Bar{z}(t) - \mathbf{H}\Bar{z}(t) + \hat{\mathbf{G}}(t)\Bar{v}_0(t)\bigr] \nonumber\\
    &\quad + \mathbf{F}_0\Bar{e}(t) - (\mathbb{L}\otimes I_n)\bigl[\mathbf{F}_0\Bar{x}(t) + \mathbf{G}_0\Bar{v}_0(t) + \Bar{d}_0(t)\bigr] \nonumber\\
    &= (\mathbb{L}\otimes I_n)\bigl[\hat{\mathbf{F}}\Bar{z}(t) + \mathbf{H}\Bar{e}(t) + \hat{\mathbf{G}}(t)\Bar{v}_0(t)\bigr] \nonumber\\
    &\quad + (\mathbb{A}_m\otimes I_n)\mathbf{F}_0\Bar{e}(t) \nonumber \\
    &\quad - (\mathbb{L}\otimes I_n)\bigl[\mathbf{F}_0\Bar{z}(t) + \mathbf{G}_0\Bar{v}_0(t) + \Bar{d}_0(t)\bigr] \nonumber\\
    &= \mathcal{H}\Bar{e}(t) + (\mathbb{L}\otimes I_n)\Xi_{\mathrm{c}}(t)\Bar{\eta}_{\mathrm{c}}(t) - \Omega\Bar{d}_0(t), \label{eq:Problem1:errordyn}
\end{align}
where
\begin{equation} \label{eq:Problem1:H}
    \mathcal{H} \coloneqq \bigl[(\mathbb{L}\otimes I_n)\mathbf{H} + (\mathbb{A}_m\otimes I_n)\mathbf{F}_0\bigr],
\end{equation}
adaptive parameter $\Xi_{\mathrm{c}}(t) \coloneqq [\Phi_{\mathrm{c}}(t), \Psi_{\mathrm{c}}(t)]$, time-varying regressor $\Bar{\eta}_{\mathrm{c}}(t) \coloneqq [\bar{z}^\top(t), \bar{v}_0^\top(t)]^\top$, and $\Omega\coloneqq(\mathbb{L}\otimes I_n)$. 

\subsection{Stability of the Coupling Operator $\mathcal{H}$}\label{subsec:Problem1:2}

The stability of \eqref{eq:Problem1:errordyn} depends on whether or not $\mathcal{H}$ in \eqref{eq:Problem1:H} is Hurwitz. We now establish sufficient conditions.
\begin{proposition}\label{prop:H_hurwitz}
    Let $\mathcal{H} \coloneqq (\mathbb{L}\otimes I_n)\mathbf{H} + (\mathbb{A}_m\otimes I_n)\mathbf{F}_0$
    where $\mathbf H=\diag\{H_1,\dots,H_m\}$ and $\mathbf F_0=I_m\otimes F_0$.  
    If there exists $P_{\mathbb L}=P_{\mathbb L}^\top\succ0$ such that
    \begin{equation*}
    \begin{aligned}
    &\mathbf H^\top(\mathbb L^\top P_{\mathbb L}\otimes I_n)
    +
    (P_{\mathbb L}\mathbb L\otimes I_n)\mathbf H  \\
    &\quad
    +
    \mathbf F_0^\top(\mathbb A_m^\top P_{\mathbb L}\otimes I_n)
    +
    (P_{\mathbb L}\mathbb A_m\otimes I_n)\mathbf F_0
    \prec 0,
    \end{aligned}
    \end{equation*}
    then $\mathcal{H}$ is Hurwitz.
\end{proposition}

The following corollary provides a more explicit robust sufficient condition in the presence of norm-bounded uncertainty in the source dynamics.

\begin{corollary}\label{cor:robust}
    Let \(F_0 = F_\ast + \Delta F\), where $\|\Delta F\|_2\le f.$
    If there exist \(P_{\mathbb L}=P_{\mathbb L}^\top\succ0\) and \(\tau>0\) such that
    \begin{equation}\label{eq:LMI_robust}
    \begin{aligned}
    &\mathbf H^\top(\mathbb L^\top P_{\mathbb L}\otimes I_n)
    +
    (P_{\mathbb L}\mathbb L\otimes I_n)\mathbf H \\
    &\quad+
    \mathbf F_\ast^\top(\mathbb A_m^\top P_{\mathbb L}\otimes I_n)
    +
    (P_{\mathbb L}\mathbb A_m\otimes I_n)\mathbf F_\ast \\
    &\quad+
    \tau I_m\otimes I_n
    \prec0,
    \end{aligned}
    \end{equation}
    and
    \begin{equation}\label{eq:LMI_robust_bound}
        2\|P_{\mathbb L}\mathbb A_m\|_2 f < \tau,
    \end{equation}
    then \(\mathcal H\) is Hurwitz for all admissible \(\Delta F\).
\end{corollary}

With the stability of the coupling operator \(\mathcal H\) ensured by Proposition~\ref{prop:H_hurwitz} and Corollary~\ref{cor:robust}, we proceed to design adaptive laws for the parameter estimates.

\subsection{Adaptive Laws and Stability}\label{subsec:Problem1:3}
We first retain the disturbance-free result (asymptotic convergence), then extend it to input-to-state stability (ISS) in the presence of bounded source disturbance.

\begin{theorem}\label{thm:stability_cont}
    Consider the networked system in \eqref{eq:Problem1:con:source}–\eqref{eq:Problem1:con:sensor}, with error dynamics \eqref{eq:Problem1:errordyn} and $\Bar{d}_0(t) = \mathbf{0}_{mn}$. 
    Let the communication graph $(\mathcal V,\mathcal E)$ satisfies the balanced and source-reachability conditions (cf. \eqref{eq:ComNet:balance} and Remark~\ref{rem:reachability}), and that the matrix $\mathcal H$ in \eqref{eq:Problem1:H} is Hurwitz. 
    Let $P_{\mathrm c}= P_{\mathrm c}^\top\succ0$ and $Q_{\mathrm c}= Q_{\mathrm c}^\top\succ0$ satisfy the Lyapunov equation
    \begin{equation}\label{eq:Lyap_Pc}
        \mathcal H^\top P_{\mathrm c} + P_{\mathrm c}\mathcal H = - Q_{\mathrm c}.
    \end{equation}
    Using $\Phi_{\mathrm c}(t)$ and $\Psi_{\mathrm c}(t)$ as in \eqref{eq:Problem1:par_error}, then the adaptive laws
    \begin{subequations}\label{eq:AdpLaw:cont}
    \begin{align}
        \dot{\Phi}_{\mathrm c}(t) &= -\Gamma_{\phi}(\mathbb{L}\otimes I_n)^\top P_{\mathrm c}\bar e(t)\bar z^\top(t),\\
        \dot{\Psi}_{\mathrm c}(t) &= -\Gamma_{\psi}(\mathbb{L}\otimes I_n)^\top P_{\mathrm c}\bar e(t)\bar v_0^\top(t),
    \end{align}
    \end{subequations}
    with gain matrices $\Gamma_{\phi},\Gamma_{\psi}\succ0$, guarantee that all closed-loop signals remain bounded and that $\bar e(t)\to0$ as $t\to\infty$.
\end{theorem}

\begin{remark}\label{rem:LaSalle}
    From LaSalle's invariance principle: the solutions of the closed-loop system  approach the largest invariant set in 
    $\mathbb{S}=\{\bar e(t)\in\mathbb{R}^{mn}:\dot V(\bar e(t),\Phi_{\mathrm c}(t),\Psi_{\mathrm c}(t))=0\}$,
    implying $\bar e(t)\to0$ and bounded $\Phi_{\mathrm c}(t),\Psi_{\mathrm c}(t)$.
\end{remark}

We now establish an ISS property when the source is subject to a bounded disturbance $\|d_{0}(t)\|\le d^{\ast}$ with the same adaptive laws \eqref{eq:AdpLaw:cont}.

\begin{theorem}\label{thm:ISS_cont}
    Consider the networked system \eqref{eq:Problem1:con:source}–\eqref{eq:Problem1:con:sensor} with error dynamics \eqref{eq:Problem1:errordyn}, $\Omega\coloneqq(\mathbb{L}\otimes I_n)$ and $\|\bar d_0(t)\|\le \bar d^\ast$ for all $t\ge0$. 
    Assume the graph satisfies \eqref{eq:ComNet:balance} and Remark~\ref{rem:reachability}, and $\mathcal H$ in \eqref{eq:Problem1:H} is Hurwitz. 
    Let $P_{\mathrm c}=P_{\mathrm c}^\top\!\succ0$ and $Q_{\mathrm c}=Q_{\mathrm c}^\top\!\succ0$ solve
    \begin{equation}\label{eq:Lyap_Pc_repeat}
        \mathcal H^\top P_{\mathrm c} + P_{\mathrm c}\mathcal H = -\,Q_{\mathrm c}.
    \end{equation}
    Apply the adaptive laws \eqref{eq:AdpLaw:cont} with $\Gamma_\phi,\Gamma_\psi\succ0$. 
    Then there exists a constant $c_{\mathrm c}(P_{\mathrm c},Q_{\mathrm c},\Omega)>0$, such that
    \begin{equation}\label{eq:ISS_bound_cont_repeat}
        \limsup_{t\to\infty}\|\bar e(t)\| \le c_{\mathrm c}\bar d^\ast.
    \end{equation}
    In particular, if $\bar d^\ast=0$, then $\bar e(t)\to0$ as $t\to\infty$.
\end{theorem}

This completes the continuous-time analysis: Theorem~\ref{thm:stability_cont} provides asymptotic convergence in the disturbance-free case, while Theorem~\ref{thm:ISS_cont} ensures ISS under bounded source disturbances. In the discrete-time setting, the behavior changes significantly—the fixed adaptive gains and normalization of regressor signals become essential to preserve stability and guarantee bounded adaptation.

\section{Adaptive Estimation for Problem 2}\label{sec:AdaptiveId}

\subsection{Error Dynamics in Networked Systems}\label{subsec:Problem2:1}
In network-level form, the discrete-time source evolves as
\begin{subequations}
\begin{align}
    \Bar{x}_{0,k+1} &= \mathbf{A}_0 \Bar{x}_{0,k} + \mathbf{B}_0 \Bar{u}_{0,k} + \Bar{\delta}_{0,k}, \label{eq:Problem2:dis:source} \\
    \Bar{y}_{0,k} &= \Bar{x}_{0,k}, \label{eq:Problem2:dis:output}
\end{align}
\end{subequations}
and the sensing node dynamics implement
\begin{subequations}
\begin{align}
    \Bar{x}_{k+1} &= \mathbf{S}\Bar{x}_{k} + (\hat{\mathbf{A}}_{k} - \mathbf{S})\Bar{z}_{k} + \hat{\mathbf{B}}_{k}\,\Bar{u}_{0,k}, \label{eq:Problem2:dis:sensor} \\
    \Bar{z}_{k} &= (\mathbb{A}_m\otimes I_n)\Bar{x}_{k} + (\mathbb{A}_0\otimes I_n)\Bar{y}_{0,k}. \label{eq:Problem2:dis:z}
\end{align}
\end{subequations}
The quantities $\Bar{x}_{0,k}$, $\Bar{u}_{0,k}$, $\Bar{\delta}_{0,k}$, and $\Bar{y}_{0,k}$ are defined via \eqref{eq:ProbFor:beta0:k}, while $\Bar{x}_{k}$ and $\Bar{z}_{k}$ use the stacking in \eqref{eq:ProbFor:betai:k}. The matrices $\mathbf{A}_0$ and $\mathbf{B}_0$ follow \eqref{eq:ProbFor:mat0} while the matrix $\mathbf{S}$ follows \eqref{eq:ProbFor:mati}. The estimate matrices $\hat{\mathbf{A}}_{k}$ and $\hat{\mathbf{B}}_{k}$ are denoted as in \eqref{eq:ProbFor:mati:k}. 

We first assume the existence of the optimal estimates.
\begin{assumption}\label{assmpt:dis:optimal}
    There exist optimal parameter values of the source dynamics in \eqref{eq:Problem2:dis:source}, denoted as $\hat{\mathbf{A}}^\ast$ and $\hat{\mathbf{B}}^\ast$, such that $\hat{\mathbf{A}}^\ast \coloneqq \mathbf{A}_0$ and $\hat{\mathbf{B}}^\ast \coloneqq \mathbf{B}_0$. 
\end{assumption}

Using 
\begin{equation}\label{eq:Problem1:con:error}
    \Bar{e}_{k} = (\mathbb{L}\otimes I_n)\Bar{x}_{k} - (\mathbb{A}_0\otimes I_n)\Bar{y}_{0,k} = \Bar{x}_{k} - \Bar{z}_{k},
\end{equation}
and introducing the parameter errors
\begin{equation}\label{eq:Problem2:par_error}
    \Phi_{\mathrm{d},k} \triangleq \hat{\mathbf{A}}_{k} - \mathbf{A}_0, \quad
    \Psi_{\mathrm{d},k} \triangleq \hat{\mathbf{B}}_{k} - \mathbf{B}_0,
\end{equation}
the estimation error dynamics, with structure analogous to \eqref{eq:Problem1:errordyn}, can be expressed as
\begin{equation}\label{eq:Problem2:errordyn}
    \Bar{e}_{k+1} = \mathcal{S}\Bar{e}_k + (\mathbb{L}\otimes I_n)\Xi_{\mathrm{d},k}\bar{\eta}_{\mathrm{d},k} - \Omega\Bar{\delta}_{0,k}.
\end{equation}
Recall that $\Omega\coloneqq(\mathbb{L}\otimes I_n)$, then
\begin{equation} \label{eq:Problem2:S}
    \mathcal{S} \coloneqq \bigl[(\mathbb{L}\otimes I_n)\mathbf{S} + (\mathbb{A}_m\otimes I_n)\mathbf{A}_0\bigr],
\end{equation}
adaptive parameter $\Xi_{\mathrm{d},k} \coloneqq [\Phi_{\mathrm{d},k}, \Psi_{\mathrm{d},k}]$ and time-varying regressor $\Bar{\eta}_{\mathrm{d},k} = [\eta_{\mathrm{d}_1,k}^\top, \dots, \eta_{\mathrm{d}_m,k}^\top]^\top \coloneqq [\bar{z}_{k}^\top, \bar{u}_{0,k}
^\top]^\top$.

\subsection{Stability of the Coupling Operator $\mathcal{S}$}\label{subsec:Problem2:2}

To guarantee exponential convergence in the discrete-time setting, we establish a sufficient condition ensuring that the coupled operator \(\mathcal S\) is Schur-stable.

\begin{proposition}\label{prop:S_schur}
    Let $\mathcal{S} \coloneqq (\mathbb{L}\otimes I_n)\mathbf{S} + (\mathbb{A}_m\otimes I_n)\mathbf{A}_0$
    where $\mathbf S=\diag\{S_1,\dots,S_m\}$ and $\mathbf A_0=I_m\otimes A_0.$
    If there exists \(P_{\mathbb L}=P_{\mathbb L}^\top\succ0\) such that
    \[
    \mathcal S^\top(P_{\mathbb L}\otimes I_n)\mathcal S
    -
    (P_{\mathbb L}\otimes I_n)
    \prec0,
    \]
    then \(\mathcal S\) is Schur-stable.
\end{proposition}

The following corollary extends this result to the case of norm-bounded uncertainty in the source dynamics, ensuring robustness of the Schur stability condition.

\begin{corollary}\label{cor:robust_S}
    Let $A_0=A_\ast+\Delta A,$ $\|\Delta A\|_2\le a,$ and define
    $ \mathbf A_\ast=I_m\otimes A_\ast$ and $\Delta\mathbf A=I_m\otimes \Delta A.$
    If there exist \(P_{\mathbb L}=P_{\mathbb L}^\top\succ0\) and \(\tau>0\) such that
    \begin{equation*}
        \mathcal S_\ast^\top(P_{\mathbb L}\otimes I_n)\mathcal S_\ast
        - (P_{\mathbb L}\otimes I_n) + \tau I_m\otimes I_n
        \prec0,
    \end{equation*}
    where \(\mathcal S_\ast = (\mathbb L\otimes I_n)\mathbf S + (\mathbb A_m\otimes I_n)\mathbf A_\ast\), 
    \(P=P_{\mathbb L}\otimes I_n\), and the uncertainty contribution satisfies the quadratic bound
    \begin{equation*}
        \Delta_{\mathcal S}^\top P\Delta_{\mathcal S}
        +
        \mathcal S_\ast^\top P\Delta_{\mathcal S}
        +
        \Delta_{\mathcal S}^\top P\mathcal S_\ast
        \preceq
        \tau I_m\otimes I_n,
    \end{equation*}
    with $\Delta_{\mathcal S} = (\mathbb A_m\otimes I_n)\Delta\mathbf A,$
    then \(\mathcal S\) is Schur-stable for all admissible \(\Delta A\).
\end{corollary}

Next, we analyze how the choice of the adaptive gains and the normalization mechanism affect performance and robustness, two aspects absent in the continuous-time case. 

\subsection{Normalized Adaptive Laws and Stability}\label{subsec:Problem2:3}
We start with a fundamental challenge. In \eqref{eq:Problem2:errordyn}, notice that the error dynamics relate the estimation and parameter errors, motivating a regression-based update for $\Xi_{\mathrm{d},k}$. 

To simplify the adaptive law design, we first consider the disturbance-free case $\bar{\delta}_{0,k}=\mathbf{0}_{mn}$, corresponding to the nominal dynamics.
In this setting, the prediction mismatch \cite{R15} can be expressed as
\begin{equation}\label{eq:Problem2:eps:def}
    \Bar{\epsilon}_{k+1} \coloneqq \Xi_{\mathrm{d},k}\Bar{\eta}_{\mathrm{d},k} = \bigl[(\mathbb{L}^\top\mathbb{L})^{-1}\mathbb{L}^\top\otimes I_n\bigr](\Bar{e}_{k+1} - \mathcal{S}\Bar{e}_k),
\end{equation}
where $\Bar{\epsilon}_{k+1} = [\epsilon_{1,k+1}^\top, \dots, \epsilon_{m,k+1}^\top ]^\top$. 
Then, the corresponding quadratic loss is
\begin{equation}\label{eq:Problem2:loss}
    \mathcal{J}_k = \tfrac{1}{2}\|\bar{\epsilon}_{k+1}\|^2
    = \tfrac{1}{2}\|\Xi_{\mathrm{d},k}\bar{\eta}_{\mathrm{d},k}\|^2,
\end{equation}
whose gradient follows directly as
\begin{equation}\label{eq:Problem2:gradloss}
    \nabla \mathcal{J}_k = \Bar{\epsilon}_{k+1}\Bar{\eta}_{\mathrm{d},k}^\top.
\end{equation}
However, the loss $\mathcal{J}_k$ is in general not strongly convex due to the time-varying regressor norm 
$\|\bar{\eta}_{\mathrm{d},k}\|^2$. Without strong convexity, \eqref{eq:Problem2:gradloss} can exhibit unbounded parameter updates. To address this and ensure a uniformly smooth optimization, we normalize the loss by 
\begin{equation}\label{eq:Problem2:normalization}
    N_k = \max\{\varphi,\, \|\bar{\eta}_{\mathrm{d},k}\|^2\}, \quad \varphi>0,
\end{equation}
(e.g., $\varphi = 0.01$) yielding the normalized cost
\begin{equation}\label{eq:Problem2:normalizedloss}
    \mathbf{J}_k = \frac{\|\Bar{\epsilon}_{k+1}\|^2}{2N_k},
    \qquad
    \nabla \mathbf{J}_k = \frac{\Bar{\epsilon}_{k+1}\Bar{\eta}_{\mathrm{d},k}^\top}{N_k}.
\end{equation}
The normalization \eqref{eq:Problem2:normalizedloss} guarantees a 1-smooth loss surface, allowing the construction of a gradient-based adaptive law. 
Since $N_k$ requires global information (not feasible for networked systems as agent $i$ does not have access to global measurements to compute $\|\bar{\eta}_{\mathrm{d},k}\|^2$), we adopt local greedy estimates
$\hat N_{i,k} = \max\{\varphi,\, \|\eta_{\mathrm{d}_i,k}\|^2\}$, leading to
\begin{equation}\label{eq:Problem2:localgradloss}
    \nabla \hat{\mathbf{J}}_k 
    = \diag\!\left\{
        \frac{\epsilon_{1,k+1}\eta_{\mathrm{d}_1,k}^\top}{\hat N_{1,k}},
        \dots,
        \frac{\epsilon_{m,k+1}\eta_{\mathrm{d}_m,k}^\top}{\hat N_{m,k}}
    \right\}.
\end{equation}
Finally, the first-order adaptive law is
\begin{equation}\label{eq:Problem2:laws}
    \hat{\Xi}_{\mathrm{d},k+1}^\top 
    = \hat{\Xi}_{\mathrm{d},k}^\top - \gamma \nabla \hat{\mathbf{J}}_k,
    \qquad \gamma>0,
\end{equation}
where $\gamma$ is a constant adaptive gain, common to all agents for simplicity of analysis.
This formulation naturally connects the distributed adaptation process with the normalized 
gradient flow, ensuring robust convergence even under non-strongly convex regressors.

Before establishing the closed-loop stability of the discrete-time network, we first analyze the parameter adaptation mechanism in \eqref{eq:Problem2:laws}. The next lemma confirms that the resulting parameter iterates remain bounded under a mild gain condition, which is fundamental for guaranteeing overall stability of the coupled error dynamics.

\begin{lemma}\label{lem:boundedness}
    Under the adaptive update \eqref{eq:Problem2:laws}, let the parameter error be defined as $\Xi_{\mathrm{d},k} \coloneqq \hat{\Xi}_{\mathrm{d},k} - \Xi_{\mathrm{d},k}^\ast$
    where $\Xi_{\mathrm{d},k}^\ast$ denotes the ideal (stationary) parameter matrix. 
    Then $\Xi_{\mathrm{d},k}$ remains bounded for all $k$ if the adaptive gain satisfies $0<\gamma<2$, 
    with $V_k = \|\Xi_{\mathrm{d},k}\|_F^2$ serving as a Lyapunov function candidate.
\end{lemma}

The range $0<\gamma<2$ guarantees a monotonic decrease of the Lyapunov energy and thus bounded parameter evolution. Having ensured that the adaptation process itself is stable, we now study the joint dynamics of the estimation error and the adaptive update in the disturbance-free case.

\begin{theorem}\label{thm:convergence}
    Consider the closed-loop network composed of \eqref{eq:Problem2:errordyn} with $\bar{\delta}_{0,k}=\mathbf{0}_{mn}$, \eqref{eq:Problem2:eps:def}, and
    the estimator update \eqref{eq:Problem2:laws}. 
    Suppose $\varphi>0$ and $0<\gamma<2$. 
    If $\mathcal{S}$ in \eqref{eq:Problem2:S} is Schur-stable, then $\lim\nolimits_{k\to\infty}\|\Bar{e}_k\| = 0$.
\end{theorem}

Theorem~\ref{thm:convergence} shows asymptotic stability in the disturbance-free case. We now reintroduce bounded process noise in the source and show that the same normalized adaptive law renders the estimation error ISS with respect to $\bar\delta_{0,k}$—the discrete-time counterpart of Theorem~\ref{thm:ISS_cont}.

\begin{theorem}\label{thm:ISS_discrete}
    Consider the networked system \eqref{eq:Problem2:dis:source}–\eqref{eq:Problem2:dis:sensor} with error dynamics \eqref{eq:Problem2:errordyn}, $\Omega\coloneqq(\mathbb L\otimes I_n)$ and $\|\bar\delta_{0,k}\|\le\bar\delta^\ast$ for all~$k$.  
    Assume $\mathcal S$ in~\eqref{eq:Problem2:S} is Schur-stable, and let $Q_{\mathrm d}=Q_{\mathrm d}^\top\succ 0$ and $P_{\mathrm d}=P_{\mathrm d}^\top\succ0$ satisfy
    \begin{equation}\label{eq:Lyap_dis}
        \mathcal S^\top P_{\mathrm d}\mathcal S - P_{\mathrm d} = -\,Q_{\mathrm d},
    \end{equation}
    Apply the normalized adaptive law~\eqref{eq:Problem2:laws} with $\gamma\in(0,2)$.  
    Then there exists a constant $c_{\mathrm d}>0$, depending only on $(P_{\mathrm d},Q_{\mathrm d},\Omega)$, such that
    \begin{equation}\label{eq:ISS_bound_dis}
        \limsup\nolimits_{k\to\infty}\|\bar e_k\|
        \le
        c_{\mathrm d}\bar\delta^\ast.
    \end{equation}
    In particular, if $\bar\delta^\ast=0$, then $\bar e_k\!\to0$ as $k\!\to\!\infty$.
\end{theorem}

In summary, the discrete-time scheme achieves (i) asymptotic convergence without disturbances and (ii) ISS under bounded process noise, with performance shaped by the Schur stability of $\mathcal S$, the constant gain $\gamma\in(0,2)$, and the per-agent normalization. This completes the discrete-time analysis and aligns it with the continuous-time guarantees.

We next validate these theoretical results through numerical simulations, demonstrating convergence of the estimation errors across network topologies and confirming scalability through computational complexity analysis as the number of sensing nodes increases.

\section{Numerical Results}\label{sec:Numerics}

\begin{figure}[t!]
    \centering
    \scalebox{0.85}{{\begin{tikzpicture}
            \centering
            % \draw[black] (-1.5,-1.5) rectangle (1.5,1.5);
            \Text[x=0,y=1.5]{$\mathcal{G}_{\mathrm{s}}\coloneqq$ Star};
            \Vertex[x=0,y=0,label=$0$,color=red,opacity=0.1,size=.5]{L}
            \Vertex[x=1,y=1,label=$1$,color=green,opacity=0.1,size=.5]{1}
            \Vertex[x=-1,y=1,label=$2$,color=green,opacity=0.1,size=.5]{2}
            \Vertex[x=-1,y=-1,label=$3$,color=green,opacity=0.1,size=.5]{3}
            \Vertex[x=1,y=-1,label=$4$,color=green,opacity=0.1,size=.5]{4}
            \Edge[Direct,color=red,label=$1.0$](L)(1)
            \Edge[Direct,color=red,label=$1.0$](L)(2)
            \Edge[Direct,color=red,label=$1.0$](L)(3)
            \Edge[Direct,color=red,label=$1.0$](L)(4)
        \end{tikzpicture}}}
    \qquad
    \scalebox{0.85}{{\begin{tikzpicture}
            \centering
            % \draw[black] (-1.5,-1.5) rectangle (1.5,1.5);
            \Text[x=0,y=1.5]{$\mathcal{G}_{\mathrm{c}}\coloneqq$ Cyclic};
            \Vertex[x=0,y=0,label=$0$,color=red,opacity=0.1,size=.5]{L}
            \Vertex[x=1,y=1,label=$1$,color=green,opacity=0.1,size=.5]{1}
            \Vertex[x=-1,y=1,label=$2$,color=green,opacity=0.1,size=.5]{2}
            \Vertex[x=-1,y=-1,label=$3$,color=green,opacity=0.1,size=.5]{3}
            \Vertex[x=1,y=-1,label=$4$,color=green,opacity=0.1,size=.5]{4}
            \Edge[Direct,color=red,label=$0.4$](L)(1)
            \Edge[Direct,color=red,label=$0.4$](L)(2)
            \Edge[Direct,color=red,label=$0.4$](L)(3)
            \Edge[Direct,color=red,label=$0.4$](L)(4)
            % \Edge[<->,label=$0.3$](1)(2)
            \draw[<->,ultra thick, latex' -latex'] (1) -- node[midway, fill=white, inner sep=2pt]{\small $0.3$} (2);
            \draw[<->,ultra thick, latex' -latex'] (2) -- node[midway, fill=white, inner sep=2pt]{\small $0.3$} (3);
            \draw[<->,ultra thick, latex' -latex'] (3) -- node[midway, fill=white, inner sep=2pt]{\small $0.3$} (4);
            \draw[<->,ultra thick, latex' -latex'] (4) -- node[midway, fill=white, inner sep=2pt]{\small $0.3$} (1);
        \end{tikzpicture}}}
    \qquad
    \scalebox{0.85}{{\begin{tikzpicture}
            \centering
            % \draw[black] (-1.5,-1.5) rectangle (1.5,1.5);
            \Text[x=0,y=1.5]{$\mathcal{G}_{\mathrm{p}}\coloneqq$ Path};
            \Vertex[x=0,y=0,label=$0$,color=red,opacity=0.1,size=.5]{L}
            \Vertex[x=1,y=1,label=$1$,color=green,opacity=0.1,size=.5]{1}
            \Vertex[x=-1,y=1,label=$2$,color=green,opacity=0.1,size=.5]{2}
            \Vertex[x=-1,y=-1,label=$3$,color=green,opacity=0.1,size=.5]{3}
            \Vertex[x=1,y=-1,label=$4$,color=green,opacity=0.1,size=.5]{4}
            \Edge[Direct,color=red,label=$1.0$](L)(1)
            \draw[<->,ultra thick, -latex'] (1) -- node[midway, fill=white, inner sep=2pt]{\small $1.0$} (2);
            \draw[<->,ultra thick, -latex'] (2) -- node[midway, fill=white, inner sep=2pt]{\small $1.0$} (3);
            \draw[<->,ultra thick, -latex'] (3) -- node[midway, fill=white, inner sep=2pt]{\small $1.0$} (4);
        \end{tikzpicture}}}
    \caption{Three network topologies $(\mathcal{G}_{\mathrm{s}}, \mathcal{G}_{\mathrm{c}}, \mathcal{G}_{\mathrm{p}})$ with weights used in the simulations.}
    \label{Fig:topology}
\end{figure}
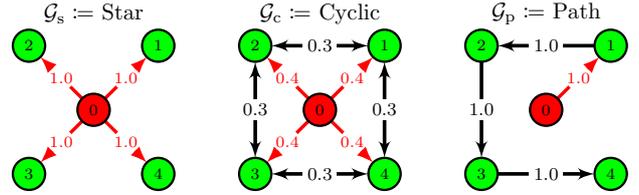

We validate the theoretical results through simulations on three network topologies—\emph{star}, \emph{cyclic}, and \emph{path} (see Fig.~\ref{Fig:topology} for details), each satisfying \eqref{eq:ComNet:balance} and Remark~\ref{rem:reachability}. For both time scales, we consider $m = 4$ sensing nodes with $n = 2$ dimensional space. For scalability tests, we extend to $m \in \{100, 150, 200, 300\}$ to measure computational complexity\footnote{ All simulations use \texttt{MATLAB R2023a} on a standard desktop. 
Continuous time: \texttt{ode45}, tolerances $10^{-6}$/ $10^{-8}$, horizon $T=30\,\mathrm{s}$. Discrete time: $K=300$ samples. For complexity, we record total CPU time via \texttt{tic}–\texttt{toc} per topology and network size.}.

\textbf{Source dynamics:} 
For continuous time,
\begin{equation*}
    F_\ast =
    \begin{bmatrix}
        0 & 1 \\
        -1 & -0.5
    \end{bmatrix}, \quad
    G_\ast =
    \begin{bmatrix}
        0 \\ 
        1
    \end{bmatrix},
\end{equation*}
with input $ v_0(t) = 0.7\sin(0.5t) + 1.5\cos(1.0t + \pi/6)$ and disturbance $d_0(t) = 5.5[\sin(0.1t),~0.5\cos(0.3t)]^\top$.
For discrete time,
\begin{equation*}
    A_\ast =
    \begin{bmatrix}
        0.9 & 0.1 \\
        -0.1 & 0.95
    \end{bmatrix}, \quad
    B_\ast =
    \begin{bmatrix}
        0.05 \\ 0.10
    \end{bmatrix},
\end{equation*}
with input $u_{0,k} = 0.9\sin(0.05k) + 0.6\cos(0.1k + \pi/5)$ and disturbance $\delta_{0,k} = 0.05[0.7\sin(0.05k),~0.5\cos(0.09k)]^\top$.
Unknowns are introduced as $\| \Delta F \|_2 = 0.55$, $\| \Delta G \|_2 = 0.50$, $\| \Delta A \|_2 = 0.1$, and $\| \Delta B \|_2 = 0.07$ to obtain $F_0$, $G_0$, $A_0$, and $B_0$ respectively.

\textbf{Sensing nodes:}
Each node implements a local observer with following details. \textit{Continuous}: $H_i = \mathrm{diag}([-2,\,-2])$ (Hurwitz), with $\Gamma_\phi = \Gamma_\psi = 10I_m$; \textit{Discrete}: $S_i = \mathrm{diag}([0.5,\,0.5])$ (Schur-stable), with $\gamma = 1.3$ and $\varphi = 0.01$.

\textbf{Initial conditions:}
The source state is initialized as $x_0(0) = [1.0,\,-0.5]^\top$ (continuous) or $x_{0,0} = [1.0,\,-0.6]^\top$ (discrete). 
Each node state satisfies $x_i(0) \in \{x \in \mathbb{R}^n : \|x\|_\infty \le 1\}$, $i=1,\dots,m$, and in simulations we draw $x_i(0)$ with components independently from $\mathrm{Unif}[-1,1]$. We design $\hat{F}_i(0)=\hat{G}_i(0)=0$ and $\hat{A}_{i,0}=\hat{B}_{i,0}=0$ as initial parameter estimates.

\begin{figure*}[t!]
    \centering
    \subfloat[\label{F3a} Star, $m=4$]{\includegraphics[width=.235\linewidth]{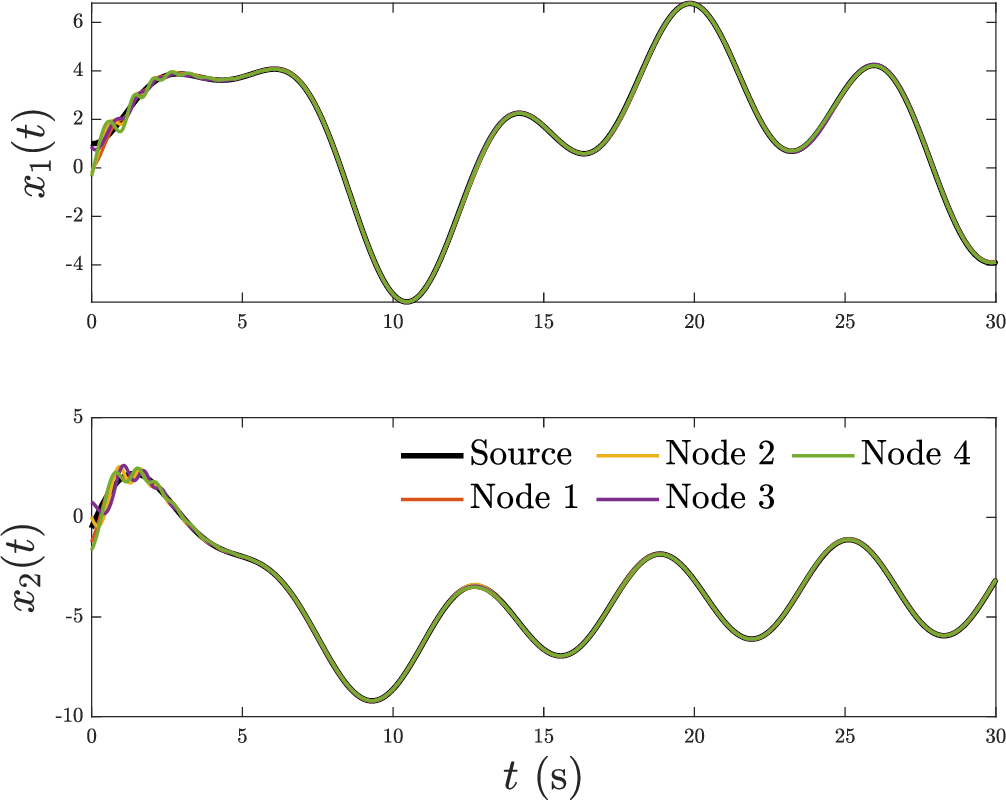}}\;
    \subfloat[\label{F3b} Cyclic, $m=4$]{\includegraphics[width=.235\linewidth]{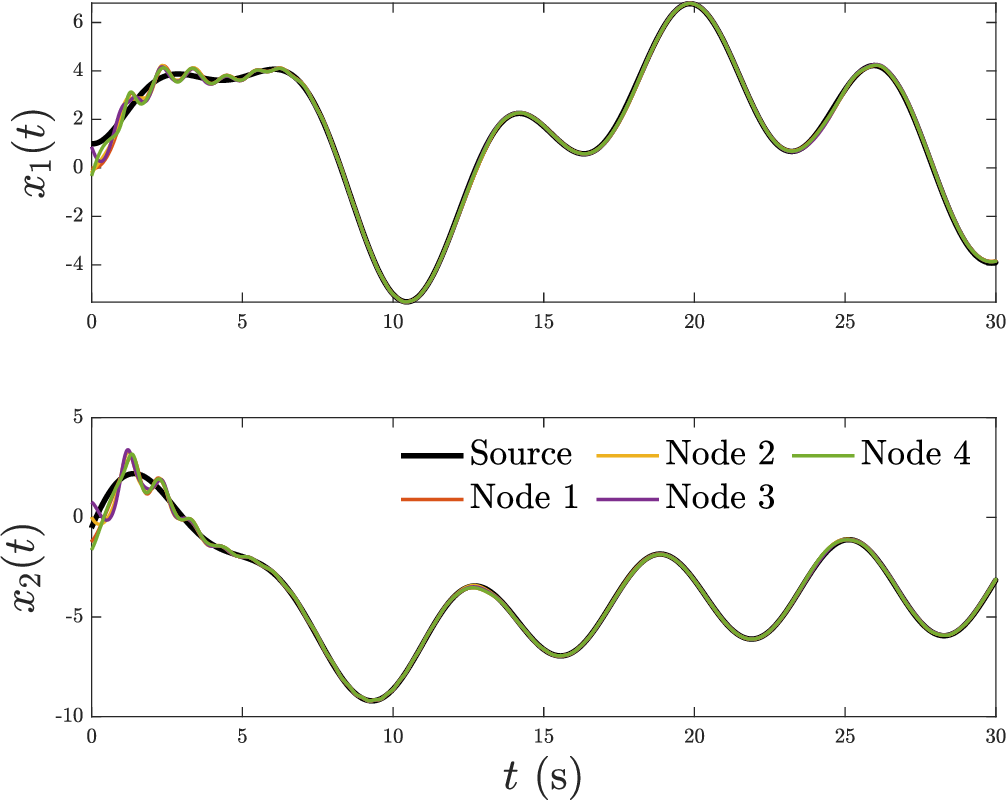}}\;
    \subfloat[\label{F3c} Path, $m=4$]{\includegraphics[width=.235\linewidth]{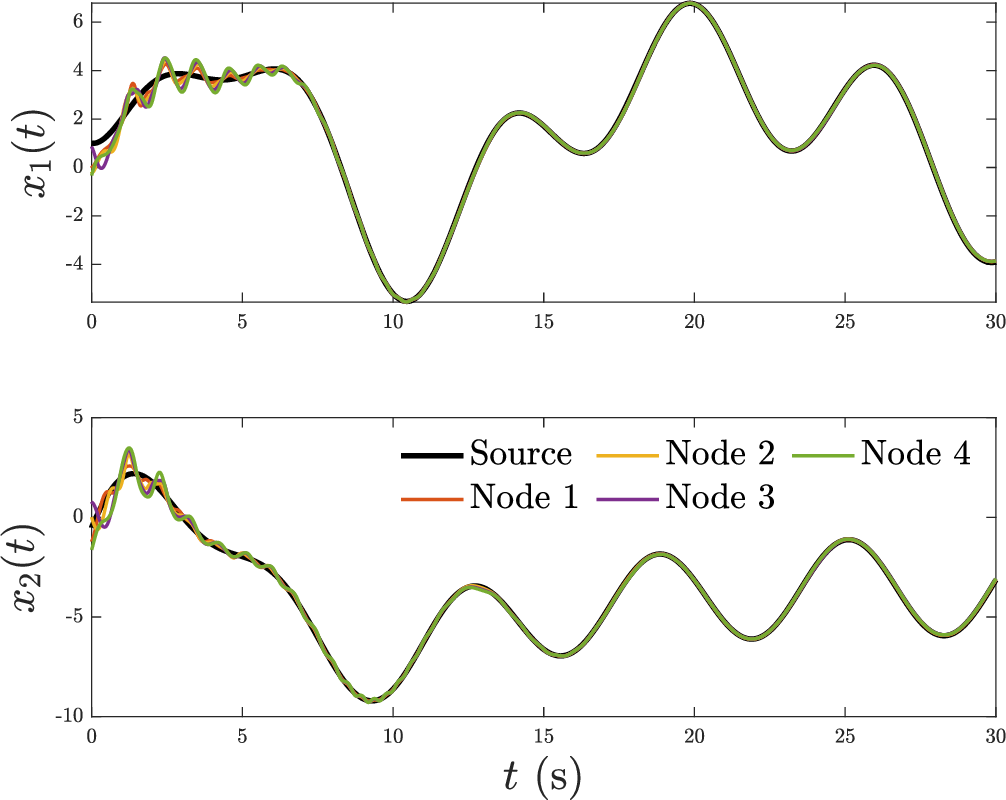}}\;
    \subfloat[\label{F3d} Error norm]{\includegraphics[width=.23\linewidth]{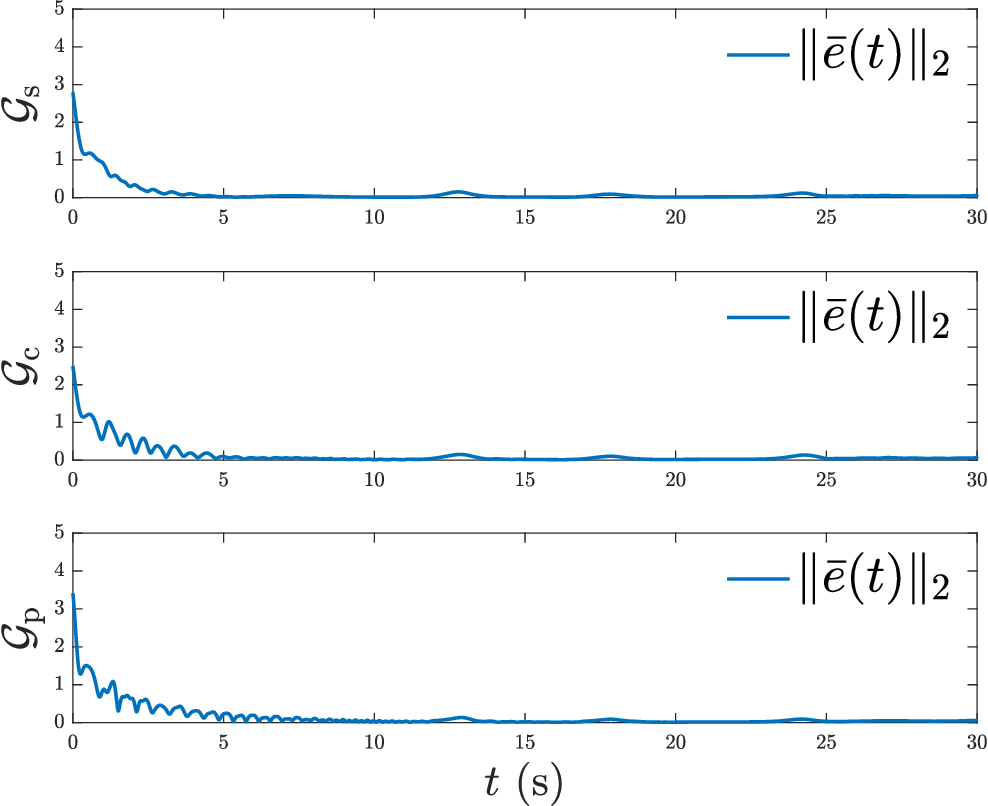}}\\
    \subfloat[\label{F3e} Star, $m=4$]{\includegraphics[width=.235\linewidth]{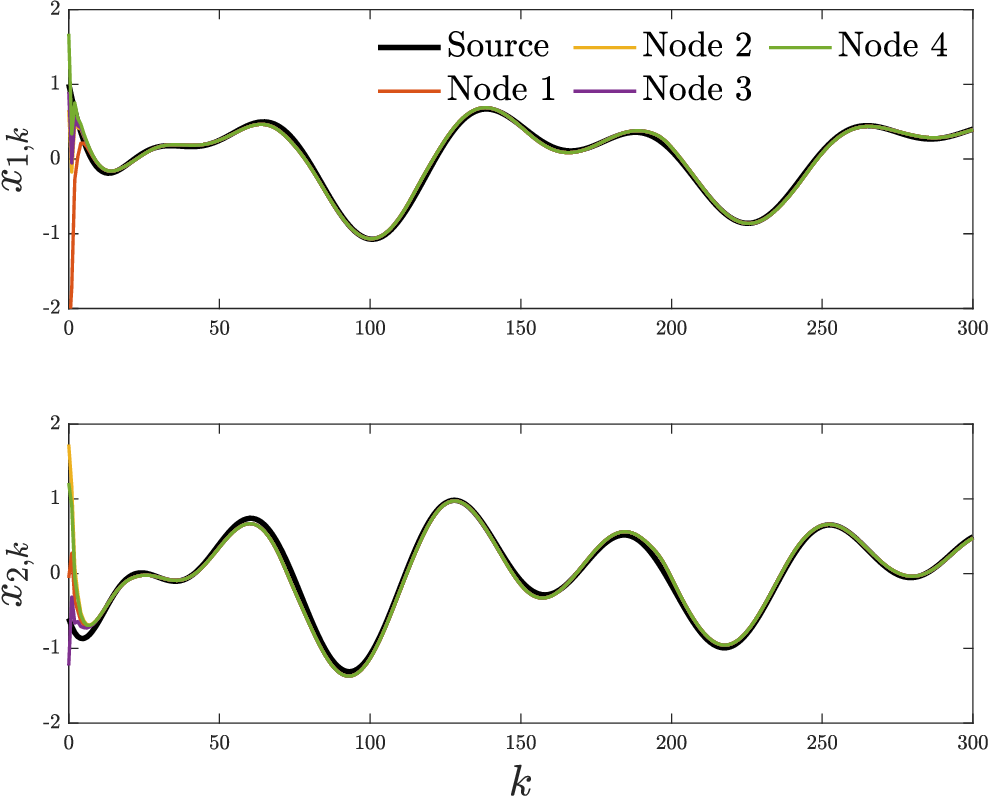}}\;
    \subfloat[\label{F3f} Cyclic, $m=4$]{\includegraphics[width=.235\linewidth]{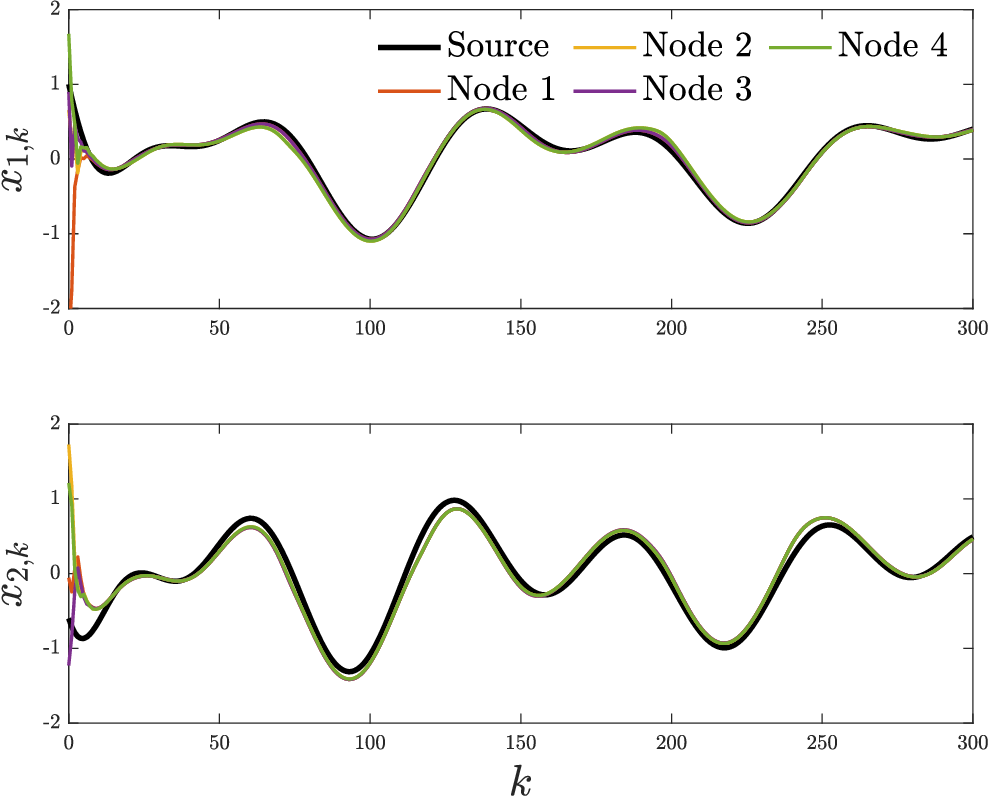}}\;
    \subfloat[\label{F3g} Path, $m=4$]{\includegraphics[width=.235\linewidth]{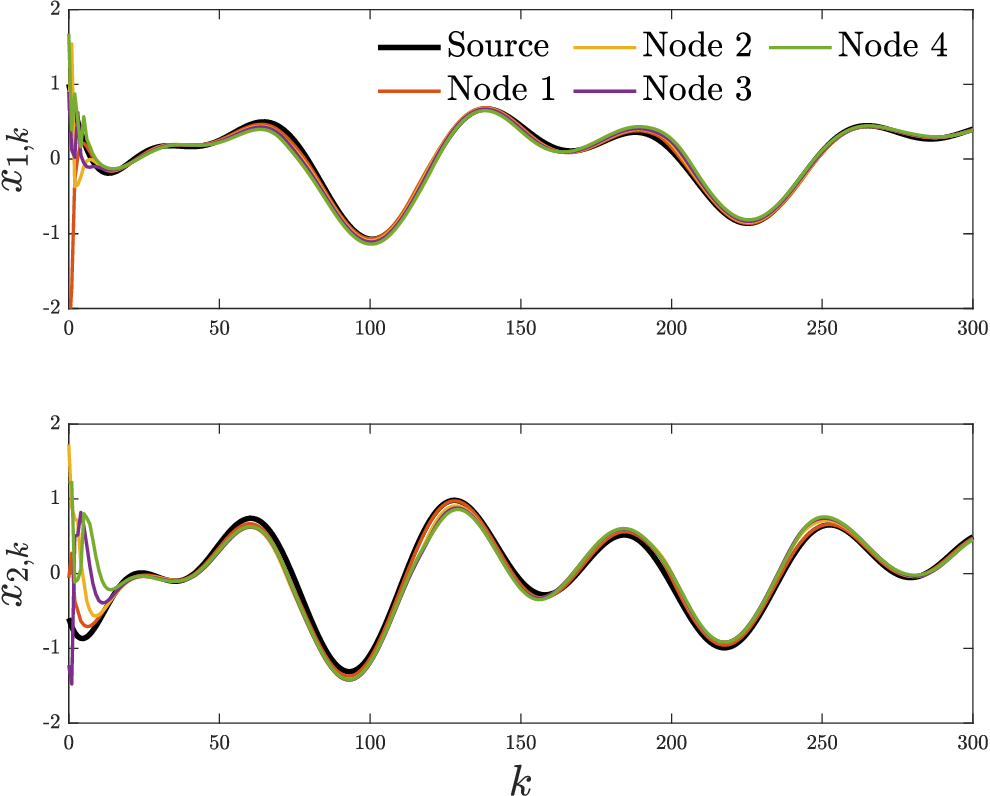}}\;
    \subfloat[\label{F3h} Error norm]{\includegraphics[width=.235\linewidth]{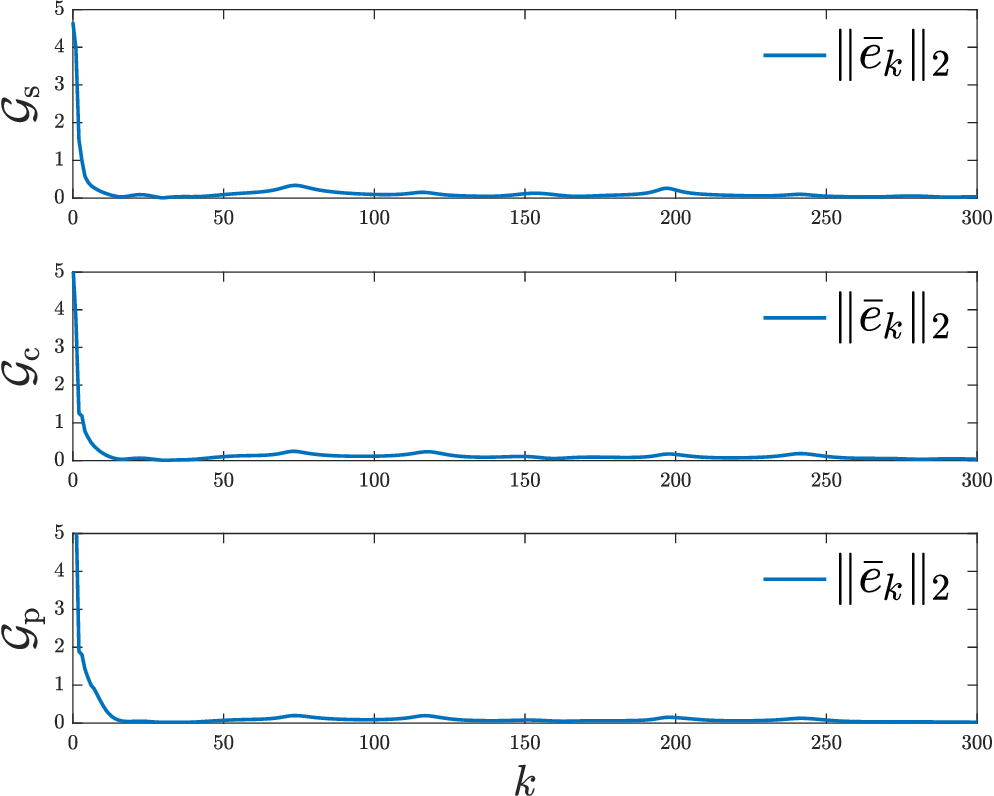}}
    \caption{Trajectories of sensing nodes for continuous-time cases (a)--(c) and discrete-time cases (e)--(g), along with corresponding error norms in (d) and (h).}
    \label{F3}
\end{figure*}

For both time scales and all three network topologies (star, cyclic, path), the trajectories of all sensing nodes closely track the source state, and the aggregated error norms decay, as illustrated in Figs.~\ref{F3a}--\ref{F3c} for continuous time and Figs.~\ref{F3e}--\ref{F3g} for discrete time, with the corresponding bounded error norms shown in Figs.~\ref{F3d} and~\ref{F3h}. The star topology performs best, which is expected since every sensing node directly receives source information. The cyclic topology uses only partial source access ($40\%$ source weighting) yet still achieves fast agreement through network coupling, demonstrating that distributed coordination can achieve near-centralized accuracy. The path topology exhibits slower convergence due to weaker connectivity and longer information flow, but still reaches a bounded error consistent with input-to-state stability (ISS). These results are consistent with the theoretical analysis: the continuous-time analysis (Proposition~\ref{prop:H_hurwitz}, Corollary~\ref{cor:robust}, Theorem~\ref{thm:ISS_cont}) guarantees stability of the network operator~$\mathcal{H}$ and ISS with respect to bounded disturbances, and the discrete-time analysis (Proposition~\ref{prop:S_schur}, Corollary~\ref{cor:robust_S}, Theorem~\ref{thm:ISS_discrete}) provides the analogous guarantees for~$\mathcal{S}$ in sampled time.

Table~\ref{tab:timing} reports the total runtime $t_{\mathrm{tot}}$ as the number of sensing nodes~$m$ increases for each topology. The discrete-time algorithm performs explicit updates with per-step complexity $\mathcal{O}(m)$, as each iteration requires only local state propagation, neighbor aggregation, and a low-rank adaptive update per node. Consequently, total runtime scales linearly with~$m$. The continuous-time implementation, based on the \texttt{ode45} solver, exhibits the same theoretical $\mathcal{O}(m)$ complexity but with larger constants due to adaptive step-size control and numerical stiffness. Among the topologies, the path graph consistently takes longer because weaker coupling slows information propagation, leading the solver to take smaller steps. The star and cyclic graphs achieve faster convergence and shorter runtimes, remaining below a few seconds even at $m{=}300$. Both continuous and discrete formulations thus demonstrate linear computational scaling and practical efficiency for large sensor networks.

Both continuous and discrete distributed adaptive estimators achieve accurate tracking, ISS robustness, and near-linear computational growth, confirming the theoretical predictions and demonstrating the framework’s scalability to large directed networks.

\begin{table}[t!]
    \centering
    \caption{Total runtime $t_{\mathrm{tot}}$ (s) vs.\ number of sensing nodes $m$ for each topology.}
    \begin{tabular}{c|ccc|c|ccc}
        \toprule
        & \multicolumn{3}{c|}{\textbf{Continuous}} & & \multicolumn{3}{c}{\textbf{Discrete}} \\
        \cmidrule(lr){2-4} \cmidrule(lr){6-8}
        $m$ & $\mathcal{G}_{\mathrm{s}}$ & $\mathcal{G}_{\mathrm{c}}$ & $\mathcal{G}_{\mathrm{p}}$ 
        & $m$ & $\mathcal{G}_{\mathrm{s}}$ & $\mathcal{G}_{\mathrm{c}}$ & $\mathcal{G}_{\mathrm{p}}$ \\
        \midrule
        100 & 0.66 & 0.65 & 1.65 & 100 & 0.45 & 0.43 & 0.38 \\
        150 & 0.76 & 0.86 & 1.95 & 150 & 0.68 & 0.69 & 0.71 \\
        200 & 0.98 & 1.23 & 2.63 & 200 & 0.85 & 0.84 & 0.84 \\
        300 & 1.88 & 2.68 & 5.38 & 300 & 1.17 & 1.18 & 1.16 \\
        \bottomrule
    \end{tabular}
    \label{tab:timing}
\end{table}

\section{Conclusion and Future Work}

We studied distributed adaptive estimation for both continuous and discrete time scales with a partially unknown source. In both cases we established stability of the network operators—$\mathcal H$ (continuous) and $\mathcal S$ (discrete)—and proved convergence of the estimation error; moreover, with bounded source disturbances we derived ISS bounds for both time scales. The main distinction is architectural: continuous-time uses smooth adaptation, whereas discrete-time requires per-step normalization with a constant gain to handle sampling and non–strongly convex regressors. Simulations across multiple topologies corroborate the theory, showing that all estimates accurately track the source despite uncertainties and disturbances. The framework also scales efficiently with the number of sensing nodes, demonstrating both robustness and computational feasibility for larger networks.

Future research will focus on extending the proposed adaptive estimation framework toward control-oriented architectures. Coupling the estimator with distributed or non-fragile controllers \cite{Wafi-Elham} can enable real-time regulation of uncertain systems using the identified source dynamics. Further directions include learning-based and geometric control formulations—such as neural-network feedback \cite{Hedesh-Robust}, reinforcement-learning adaptation under packet loss \cite{Wafi-SIAM}, and convex data-driven control with Riemannian metrics [\cite{R16}]. Integrating safety principles and density-based coordination \cite{R17}, as well as robust active control in network systems \cite{R18,Wafi-ASCC26-DNC,Wafi-ASCC26-FTC}, strengthen the connection between adaptive estimation, control, and safe multi-agent interaction.

\bibliographystyle{ieeetr}
\bibliography{EN-Bib.bib}

\appendix

\section{Proofs}
    % \subsection*{Proof of Proposition~\ref{prop:H_hurwitz}}
    \textbf{Proof of Proposition~\ref{prop:H_hurwitz}:} 
    Let $P=P_{\mathbb L}\otimes I_n\succ0$ and consider $V(\bar e)=\bar e^\top P\bar e.$
    Along $\dot{\bar e}=\mathcal H\bar e,$ we have
    \begin{equation*}
        \dot V = \bar e^\top(\mathcal H^\top P+P\mathcal H)\bar e.
    \end{equation*}
    Using $\mathcal{H} \coloneqq (\mathbb{L}\otimes I_n)\mathbf{H} + (\mathbb{A}_m\otimes I_n)\mathbf{F}_0$,
    we obtain
    \begin{equation*}
        \begin{aligned}
        \mathcal H^\top P+P\mathcal H
        &=
        \mathbf H^\top(\mathbb L^\top P_{\mathbb L}\otimes I_n)
        +
        (P_{\mathbb L}\mathbb L\otimes I_n)\mathbf H \\
        &\quad+
        \mathbf F_0^\top(\mathbb A_m^\top P_{\mathbb L}\otimes I_n)
        +
        (P_{\mathbb L}\mathbb A_m\otimes I_n)\mathbf F_0.
        \end{aligned}
    \end{equation*}
    By assumption, this matrix is negative definite. Hence $\dot V<0, \forall \bar e\neq0.$
    Therefore, $\mathcal H$ is Hurwitz.
    
    % \subsection*{Proof of Corollary~\ref{cor:robust}}
    \textbf{Proof of Corollary~\ref{cor:robust}:}
    Let $P=P_{\mathbb L}\otimes I_n.$ By definition \(F_0=F_\ast+\Delta F\), define
    $\mathbf F_\ast=I_m\otimes F_\ast$ and $\Delta\mathbf F=I_m\otimes \Delta F$.
    Then $\mathbf F_0=\mathbf F_\ast+\Delta\mathbf F.$
    Using Proposition~\ref{prop:H_hurwitz}, it is sufficient to show that
    $\mathcal H^\top P+P\mathcal H\prec0$ for all admissible \(\Delta F\). Expanding the \(F_0\)-dependent part gives
    \begin{equation*}
        \begin{aligned}
        \mathcal H^\top P+P\mathcal H
        &=
        \mathbf H^\top(\mathbb L^\top P_{\mathbb L}\otimes I_n)
        +
        (P_{\mathbb L}\mathbb L\otimes I_n)\mathbf H \\
        &\quad+
        \mathbf F_\ast^\top(\mathbb A_m^\top P_{\mathbb L}\otimes I_n)
        +
        (P_{\mathbb L}\mathbb A_m\otimes I_n)\mathbf F_\ast \\
        &\quad+
        \Delta\mathbf F^\top(\mathbb A_m^\top P_{\mathbb L}\otimes I_n)
        +
        (P_{\mathbb L}\mathbb A_m\otimes I_n)\Delta\mathbf F .
        \end{aligned}
    \end{equation*}
    The uncertain part satisfies, for any \(\bar e\),
    \begin{equation*}
        \begin{aligned}
        &\bar e^\top
        \left[
        \Delta\mathbf F^\top(\mathbb A_m^\top P_{\mathbb L}\otimes I_n)
        +
        (P_{\mathbb L}\mathbb A_m\otimes I_n)\Delta\mathbf F
        \right]\bar e \\
        &\qquad
        \le
        2\|P_{\mathbb L}\mathbb A_m\|_2
        \|\Delta F\|_2
        \|\bar e\|^2 \\
        &\qquad
        \le
        2\|P_{\mathbb L}\mathbb A_m\|_2 f\|\bar e\|^2.
        \end{aligned}
    \end{equation*}
    By \eqref{eq:LMI_robust_bound}, $2\|P_{\mathbb L}\mathbb A_m\|_2 f<\tau.$
    Therefore, the uncertain contribution is dominated by $\tau I_m\otimes I_n.$
    Since \eqref{eq:LMI_robust} states that the nominal part plus this slack term is negative definite, it follows that $\mathcal H^\top P+P\mathcal H\prec0$
    for all admissible \(\Delta F\). Hence, by Proposition~\ref{prop:H_hurwitz}, \(\mathcal H\) is Hurwitz.

    % \subsection*{Proof of Theorem~\ref{thm:stability_cont}}
    \textbf{Proof of Theorem~\ref{thm:stability_cont}:}
        Define the Lyapunov function
        \begin{equation}\label{eq:Lyap_V}
            \begin{aligned}
                V(t) & = \Bar{e}^\top(t)\mathbf{P}_{\mathrm c}\Bar{e}(t) + \tr\bigl[\Phi_{\mathrm c}^\top(t)\Gamma_\phi^{-1}\Phi_{\mathrm c}(t)\bigr] \\
                & \quad + \tr\bigl[\Psi_{\mathrm c}^\top(t)\Gamma_\psi^{-1}\Psi_{\mathrm c}(t)\bigr]. 
            \end{aligned}
        \end{equation}
        Differentiating \eqref{eq:Lyap_V} along \eqref{eq:Problem1:errordyn} and using \eqref{eq:Lyap_Pc} gives
        \begin{align*}
            \dot V(t)
            &= \bar e^\top(t)(\mathcal H^\top P_{\mathrm c} + P_{\mathrm c}\mathcal H)\bar e(t) \\
            &\quad +2\bar e^\top(t) P_{\mathrm c}(\mathbb{L}\otimes I_n)
                  \bigl[\Phi_{\mathrm c}(t)\bar z(t) + \Psi_{\mathrm c}(t)\bar v_0(t)\bigr]\\
            &\quad +2\tr\bigl[\dot\Phi_{\mathrm c}^\top(t)\Gamma_{\phi}^{-1}\Phi_{\mathrm c}(t)\bigr]
                   +2\tr\bigl[\dot\Psi_{\mathrm c}^\top(t)\Gamma_{\psi}^{-1}\Psi_{\mathrm c}(t)\bigr].
        \end{align*}
        Substituting \eqref{eq:AdpLaw:cont} yields perfect cancellation of the cross terms, leading to
        \begin{equation*}
            \dot V(t)=-\bar e(t)^\top Q_{\mathrm c}\bar e(t) \le 0.
        \end{equation*}
        Therefore, the Lyapunov function $V(t)$ is nonincreasing and all trajectories are bounded.  
        Because $\dot V(t)=0$ if and only if $\bar e(t)=0$, the largest invariant set contained in $\{\dot V(t)=0\}$ corresponds to $\bar e(t)\equiv0$, $\dot{\Phi}_{\mathrm c}=\dot{\Psi}_{\mathrm c}=0$.  
        By LaSalle’s invariance principle, $\bar e(t)\to0$ and the adaptive parameters converge to constant steady-state values.  

    % \subsection*{Proof of Theorem~\ref{thm:ISS_cont}}
    \textbf{Proof of Theorem~\ref{thm:ISS_cont}:}
        Consider the Lyapunov function $V(t)$ given in \eqref{eq:Lyap_V} with $\Phi_{\mathrm c}(t),\Psi_{\mathrm c}(t)$ as in \eqref{eq:Problem1:par_error}.
        Along the error dynamics and using the adaptive laws \eqref{eq:AdpLaw:cont}, all mixed terms involving $\Xi_{\mathrm c}(t)\bar\eta_{\mathrm c}(t)$ cancel by construction. Using \eqref{eq:Lyap_Pc_repeat} we obtain
        \begin{equation*}
            \dot V(t) = -\bar e(t)^\top Q_{\mathrm c}\bar e(t)
            + 2\bar e(t)^\top P_{\mathrm c}\Omega\bar d_0(t).
        \end{equation*}
        For any $\varepsilon\in(0,1)$, by Cauchy–Schwarz and completing the square
        \begin{equation*}\begin{aligned}
            2\bar e(t)^\top P_{\mathrm c}\Omega\bar d_0(t)
            &\le \varepsilon\bar e(t)^\top Q_{\mathrm c}\bar e(t) \\
            &\quad + \tfrac{1}{\varepsilon}
            \lambda_{\max}\big(P_{\mathrm c}\Omega\Omega^\top P_{\mathrm c}\big)\|\bar d_0(t)\|^2.
            \end{aligned}
        \end{equation*}
        Hence
        \begin{equation*}\begin{aligned}
            \dot V(t) &\le 
            -(1-\varepsilon)\bar e(t)^\top Q_{\mathrm c}\bar e(t) \\
            &\quad + \tfrac{1}{\varepsilon}\,
            \lambda_{\max}\!\big(P_{\mathrm c}\Omega\Omega^\top P_{\mathrm c}\big)\,\|\bar d_0(t)\|^2.
            \end{aligned}
        \end{equation*}
        Using $\bar e(t)^\top Q_{\mathrm c}\bar e(t) \ge \lambda_{\min}(Q_{\mathrm c})\|\bar e(t)\|^2$ and 
        $\bar e(t)^\top P_{\mathrm c}\bar e(t) \ge \lambda_{\min}(P_{\mathrm c})\|\bar e(t)\|^2$, we get
        \begin{equation*}
            \dot V(t) \le -\alpha_{\mathrm c}\|\bar e(t)\|^2 + \beta_{\mathrm c}\|\bar d_0(t)\|^2,
        \end{equation*}
        where $\alpha_{\mathrm c} \coloneqq (1-\varepsilon)\lambda_{\min}(Q_{\mathrm c})$ and $\beta_{\mathrm c} \coloneqq \frac{1}{\varepsilon}\lambda_{\max}(P_{\mathrm c}\Omega\Omega^\top P_{\mathrm c})$.
        By standard ISS comparison arguments, it follows that
        \begin{equation*}
            \limsup_{t\to\infty}\|\bar e(t)\|
            \le
            \sqrt{\frac{\beta_{\mathrm c}}{\alpha_{\mathrm c}\lambda_{\min}(P_{\mathrm c})}}
            \;\bar d^\ast =: c_{\mathrm c}\bar d^\ast,
        \end{equation*}
        which establishes \eqref{eq:ISS_bound_cont_repeat}. If $\bar d^\ast=0$, then $\dot V(t)\le -\alpha\|\bar e(t)\|^2$ and thus $\bar e(t)\to0$.
    
    % \subsection*{Proof of Proposition~\ref{prop:S_schur}}
    \textbf{Proof of Proposition~\ref{prop:S_schur}:}
        Let \(P=P_{\mathbb L}\otimes I_n\succ0\) and consider $V_k=\bar e_k^\top P\bar e_k.$
        Along \(\bar e_{k+1}=\mathcal S\bar e_k\),
        \begin{equation*}
            V_{k+1}-V_k
            =
            \bar e_k^\top
            \left(
            \mathcal S^\top P\mathcal S-P
            \right)
            \bar e_k.
        \end{equation*}
        By assumption, $\mathcal S^\top P\mathcal S-P\prec0.$
        Hence \(V_{k+1}-V_k<0\) for all \(\bar e_k\neq0\), and therefore \(\mathcal S\) is Schur-stable.

    % \subsection*{Proof of Corollary~\ref{cor:robust_S}}
    \textbf{Proof of Corollary~\ref{cor:robust_S}:}
        Let $P=P_{\mathbb L}\otimes I_n.$ Since $A_0=A_\ast+\Delta A,$ we have
        $\mathbf A_0=\mathbf A_\ast+\Delta\mathbf A$ and $\Delta\mathbf A=I_m\otimes \Delta A.$
        Hence $\mathcal S = \mathcal S_\ast+\Delta_{\mathcal S},$ where
        \begin{equation*}
            \mathcal S_\ast
            = (\mathbb L\otimes I_n)\mathbf S + (\mathbb A_m\otimes I_n)\mathbf A_\ast,
        \end{equation*}
        and
        \begin{equation*}
            \Delta_{\mathcal S} = (\mathbb A_m\otimes I_n)\Delta\mathbf A.
        \end{equation*}
        By Proposition~\ref{prop:S_schur}, it is sufficient to show that
        \begin{equation*}
            \mathcal S^\top P\mathcal S-P\prec0.
        \end{equation*}
        Expanding gives
        \begin{equation*}
            \begin{aligned}
            \mathcal S^\top P\mathcal S-P
            &=
            \mathcal S_\ast^\top P\mathcal S_\ast-P \\
            &\quad+
            \mathcal S_\ast^\top P\Delta_{\mathcal S}
            +
            \Delta_{\mathcal S}^\top P\mathcal S_\ast
            +
            \Delta_{\mathcal S}^\top P\Delta_{\mathcal S}.
            \end{aligned}
        \end{equation*}
        By assumption,
        \begin{equation*}
            \mathcal S_\ast^\top P\mathcal S_\ast-P
            + \tau I_m\otimes I_n \prec0.
        \end{equation*}
        Also, the uncertainty contribution satisfies
        \begin{equation*}
            \Delta_{\mathcal S}^\top P\Delta_{\mathcal S}
            +
            \mathcal S_\ast^\top P\Delta_{\mathcal S}
            +
            \Delta_{\mathcal S}^\top P\mathcal S_\ast
            \preceq
            \tau I_m\otimes I_n.
        \end{equation*}
        Therefore, $\mathcal S^\top P\mathcal S-P\prec0.$
        Hence, by Proposition~\ref{prop:S_schur}, \(\mathcal S\) is Schur-stable for all admissible 
        \(\Delta A\).

    % \subsection*{Proof of Lemma~\ref{lem:boundedness}}
    \textbf{Proof of Lemma~\ref{lem:boundedness}:}
        Define the Lyapunov function \( V_k = \|\Xi_{\mathrm{d},k}\|_F^2 = \tr(\Xi_{\mathrm{d},k}^\top \Xi_{\mathrm{d},k}) \).  
        Using the adaptive update \eqref{eq:Problem2:laws}, the parameter recursion can be written as
        \begin{equation*}
            \Xi_{\mathrm{d},k+1} = \Xi_{\mathrm{d},k} - \gamma \nabla \hat{\mathbf{J}}_k,
        \end{equation*}
        where the local gradient is 
        \begin{equation*}
            \nabla \hat{\mathbf{J}}_k = 
            \diag\left\{
                \frac{\epsilon_{1,k+1}\eta_{\mathrm{d}_1,k}^\top}{\hat{N}_{1,k}}, \dots, 
                \frac{\epsilon_{m,k+1}\eta_{\mathrm{d}_m,k}^\top}{\hat{N}_{m,k}}
            \right\}.
        \end{equation*}
        Then, the one-step difference of $V_k$ is
        \begin{align}
            \Delta V_k \coloneqq V_{k+1} - V_k 
            &= \|\Xi_{\mathrm{d},k+1}\|_F^2 - \|\Xi_{\mathrm{d},k}\|_F^2 \nonumber\\
            &= \|\Xi_{\mathrm{d},k} - \gamma \nabla \hat{\mathbf{J}}_k\|_F^2 - \|\Xi_{\mathrm{d},k}\|_F^2. \label{eq:lemma:Vk_diff}
        \end{align}
        Expanding \eqref{eq:lemma:Vk_diff} and rearranging gives
        \begin{align}
            \Delta V_k 
            &= -2\gamma\tr\big(\Xi_{\mathrm{d},k}^\top \nabla \hat{\mathbf{J}}_k\big)
              + \gamma^2 \|\nabla \hat{\mathbf{J}}_k\|_F^2. \label{eq:lemma:Vk_expand}
        \end{align}
        From the definition of $\nabla\hat{\mathbf{J}}_k$, we have
        \begin{equation*}
            \tr\big(\Xi_{\mathrm{d},k}^\top \nabla \hat{\mathbf{J}}_k\big)
            = \sum_{i=1}^{m} \frac{1}{\hat N_{i,k}} 
              \epsilon_{i,k+1}^\top \Xi_{\mathrm{d}_i,k} \eta_{\mathrm{d}_i,k}.
        \end{equation*}
        Using the regression form $\epsilon_{i,k+1} = \Xi_{\mathrm{d}_i,k}\eta_{\mathrm{d}_i,k}$ and 
        the fact that $\hat N_{i,k} \ge \|\eta_{\mathrm{d}_i,k}\|^2$, it follows that
        \begin{equation*}
            \Delta V_{i,k} = 
            -\frac{\gamma}{\hat N_{i,k}}\!\left(2 - \frac{\gamma \|\eta_{\mathrm{d}_i,k}\|^2}{\hat N_{i,k}}\right)\!
            \|\epsilon_{i,k+1}\|^2 \le 0,
        \end{equation*}
        for $0<\gamma<2$ and thus $\Delta V_k = \sum_i \Delta V_{i,k}\le 0$ globally. 
        Hence $V_k$ is nonincreasing and bounded below by zero, implying bounded $\Xi_{\mathrm{d},k}$ for all $k$. 

    % \subsection*{Proof of Theorem~\ref{thm:convergence}}
    \textbf{Proof of Theorem~\ref{thm:convergence}:}
        From Lemma~\ref{lem:boundedness}, $\Xi_{\mathrm{d},k}$ is bounded and $V_k$ nonincreasing. 
        Summing \eqref{eq:lemma:Vk_expand} over $k$ gives
        \begin{equation*}
            \sum_{k=0}^{\infty} 
            \frac{\|\Bar{\epsilon}_{k+1}\|^2}{N_k} 
            \le \frac{V_0}{\gamma(2-\gamma)} < \infty,
        \end{equation*}
        which implies $\|\Bar{\epsilon}_{k+1}\|/N_k \to 0$ as $k\to\infty$. 
        With bounded regressors $\|\Bar{\eta}_{\mathrm{d},k}\|^2$, it follows that 
        $\Bar{\epsilon}_{k}\to\!0$.  
        
        Substituting $\Bar{\epsilon}_k$ into the error recursion \eqref{eq:Problem2:errordyn} yields
        \begin{equation*}
            \Bar{e}_{k+1} 
            = \mathcal{S}\Bar{e}_k 
              + (\mathbb{L}\otimes I_n)\Bar{\epsilon}_{k+1}.
        \end{equation*}
        Since $\mathcal{S}$ is Schur-stable, there exist constants $c>0$ and $\alpha\in(0,1)$ such that 
        $\|\mathcal{S}^\ell\|\le c\alpha^\ell$ for all $\ell\ge 0$. 
        Unfolding the recursion gives
        \begin{equation*}
            \Bar{e}_k 
            = \mathcal{S}^{k-k_0}\Bar{e}_{k_0} 
              + \sum_{t=k_0}^{k-1} 
                \mathcal{S}^{k-1-t} (\mathbb{L}\otimes I_n)\Bar{\epsilon}_{t+1}.
        \end{equation*}
        The first term decays exponentially, and the second term is a convolution 
        of an exponentially decaying kernel with a vanishing sequence $\Bar{\epsilon}_{t+1}\!\to\! 0$, 
        hence also tends to zero. 
        Then, $\lim_{k\to\infty}\|\Bar{e}_k\| = 0$, showing asymptotic convergence.

    % \subsection*{Proof of Theorem~\ref{thm:convergence}}
    \textbf{Proof of Theorem~\ref{thm:ISS_discrete}:}
        Define the Lyapunov function
        \begin{equation*}
            V_k =
            \bar e_k^\top P_{\mathrm d}\bar e_k
            +
            \tr\!\big(\Phi_{\mathrm d,k}^\top\Gamma_\phi^{-1}\Phi_{\mathrm d,k}\big)
            +
            \tr\!\big(\Psi_{\mathrm d,k}^\top\Gamma_\psi^{-1}\Psi_{\mathrm d,k}\big).
        \end{equation*}
        Under the update~\eqref{eq:Problem2:laws}, the cross-terms in~$\Xi_{\mathrm d,k}\bar\eta_{\mathrm d,k}$ cancel, and from~\eqref{eq:Lyap_dis},
        \begin{align*}
            V_{k+1}-V_k
            &\le
            -\bar e_k^\top Q_{\mathrm d}\bar e_k
            +
            2\bar e_k^\top P_{\mathrm d}\Omega\bar\delta_{0,k}.
        \end{align*}
        Applying the discrete-time inequality
        \begin{equation*}
            2x_k^\top Yz_k\le\varepsilon x_k^\top Q_{\mathrm d}x_k + \tfrac{1}{\varepsilon}\lambda_{\max}(YQ_{\mathrm d}^{-1}Y^\top)\|z_k\|^2,
        \end{equation*}
        for any $\varepsilon\in(0,1)$ gives
        \begin{equation*}\begin{aligned}
            V_{k+1}-V_k
            &\le
            -(1-\varepsilon)\,\bar e_k^\top Q_{\mathrm d}\bar e_k \\
            &\quad +
            \tfrac{1}{\varepsilon}\,
            \lambda_{\max}\big(P_{\mathrm d}\Omega Q_{\mathrm d}^{-1}\Omega^\top P_{\mathrm d}\big)\|\bar\delta_{0,k}\|^2.
            \end{aligned}
        \end{equation*}
        Using $\bar e_k^\top Q_{\mathrm d}\bar e_k \ge \lambda_{\min}(Q_{\mathrm d})\|\bar e_k\|^2$ and
        $V_k\ge\lambda_{\min}(P_{\mathrm d})\|\bar e_k\|^2$, 
        we obtain a standard ISS recursion
        \begin{equation*}
            V_{k+1}
            \le
            (1-\alpha_{\mathrm d})V_k
            +
            \beta_{\mathrm d}\|\bar\delta_{0,k}\|^2,
        \end{equation*}
        for some $\alpha_{\mathrm d},\beta_{\mathrm d}>0$.  
        By the discrete-time comparison lemma, this implies
        \begin{equation*}
            \limsup_{k\to\infty}\|\bar e_k\|
            \le
            \sqrt{\frac{\beta_{\mathrm d}}{\alpha_{\mathrm d}\lambda_{\min}(P_{\mathrm d})}}\,
           \bar\delta^\ast
           =: c_{\mathrm d}\bar\delta^\ast,
        \end{equation*}
        proving~\eqref{eq:ISS_bound_dis}.  When $\bar\delta^\ast=0$, the negative-definite term dominates and $\bar e_k\!\to\!0$.

\end{document}